\documentclass[12pt]{article}

\usepackage{filecontents}
\usepackage{amssymb, amsmath, bm, bbm, nicefrac,hyperref,float}
\usepackage{graphicx, subcaption, booktabs, enumitem,verbatim, cleveref}
\usepackage{setspace, lineno}
\usepackage[authoryear]{natbib}
\usepackage[dvipsnames]{xcolor}
\usepackage{xr}
\makeatletter
\newcommand*{\addFileDependency}[1]{
  \typeout{(#1)}
  \@addtofilelist{#1}
  \IfFileExists{#1}{}{\typeout{No file #1.}}
}
\makeatother

\newcommand*{\myexternaldocument}[1]{%
    \externaldocument{#1}%
    \addFileDependency{#1.tex}%
    \addFileDependency{#1.aux}%
}
\myexternaldocument{supplementary}

\allowdisplaybreaks  
\newcommand{\bbeta}{\bm{\beta}}

\newtheorem{theorem}{Theorem}
\newtheorem{assumption}{Assumption}

\newcommand{\blind}{1}

\begin{document}

\def\spacingset#1{\renewcommand{\baselinestretch}%
{#1}\small\normalsize} \spacingset{1}


\if1\blind
{
  \title{\bf Matching Estimators of Causal Effects in Clustered Observational Studies with Application to Quantifying the Impact of Marine Protected Areas on Biodiversity}
  \author{Can Cui, \hspace{.2cm}Shu Yang, \hspace{.2cm}Brian J Reich\hspace{.2cm}\\
    Department of Statistics, North Carolina State University\\
    and \\
    David A Gill \\
    Nicholas School of the Environment, Duke University}
    \date{}
  \maketitle
} \fi

\if0\blind
{
  \bigskip
  \bigskip
  \bigskip
  \begin{center}
    {\LARGE\bf Matching Estimators of Causal Effects in Clustered Observational Studies with Application to Quantifying the Impact of Marine Protected Areas on Biodiversity}
\end{center}
  \medskip
} \fi

\bigskip


\begin{abstract}
Marine conservation preserves fish biodiversity, protects marine and coastal ecosystems, and supports climate resilience and adaptation. Despite the importance of establishing marine protected areas (MPAs), research on the effectiveness of MPAs with different conservation policies is limited due to the lack of quantitative MPA information. In this paper, leveraging a global MPA database, we investigate the causal impact of MPA policies on fish biodiversity. To address challenges posed by this clustered and confounded observational study, we construct a matching estimator of the average treatment effect and a cluster-weighted bootstrap method for variance estimation. We establish the theoretical guarantees of the matching estimator and its variance estimator. Under our proposed matching framework, we recommend matching on both cluster-level and unit-level covariates to achieve efficiency. The simulation results demonstrate that our matching strategy minimizes the bias and achieves the nominal confidence interval coverage. Applying our proposed matching method to compare different MPA policies reveals that the no-take policy is more effective than the multi-use policy in preserving fish biodiversity. 
\end{abstract}

\vfill

\newpage
\singlespacing

\section{Introduction} \label{s:intro}
\subsection{Causal Impact of Marine Protected Areas on Biodiversity}
Preserving marine biological diversity is an important objective of governments, scientists, local communities, and conservationists. Marine protected areas (MPAs) have been established worldwide to keep sustainable and resilient marine ecosystems by restricting destructive and extractive activities within their boundaries \citep{Grorud:2021, UNEP:2021}. Despite widespread use, the effectiveness of many MPAs and different types of MPA policies in conserving marine biodiversity remain unclear \citep{Grorud:2021}. Very few studies employ rigorous causal inference methods to assess MPA impacts, and even less so to investigate the relative effects of different conservation policies \citep{Ferraro:2019}. Such studies, however, are important and have significant policy implications, as prohibiting fishing activities that are potentially important for local food and livelihood security can result in significant social costs and harm (e.g., \citet{Kamat:2014, Bennett:2014}). 

\citet{Gill:2017} investigated the effectiveness of MPA management and its impacts on fish populations. They developed a database of ecological, management, social, and environmental conditions in and around hundreds of MPAs globally. In their study, management attributes such as available capacity were strongly associated with increases in fish biomass observed in MPAs. Nonetheless, the relative effects of different types of MPAs (referred to as policies or treatments), such as those that restrict fishing (hereafter called multi-use or MU MPAs) and those that prohibit all fishing (hereafter called no-take or NT MPAs) require further investigation. 


While the \citet{Gill:2017} database represents one of the largest global datasets of MPA conditions and ecological outcomes to date, its properties present significant challenges for applying traditional causal inference methods. First, given the intractability of conducting randomized experiments in many conservation settings, the global MPA dataset is observational, and thus subject to confounding biases not present when treatment is randomized \citep{Pynegar:2021}. MU and NT MPAs are likely to be located in areas with different social, environmental and regulatory conditions. Direct comparisons of the biodiversity between MU and NT MPAs are fallible. 
Second, the MPA data are spatially clustered as nearby sites are usually under the same conservation policy, whether it be because they lie within the same MPA, specific management zone within an MPA (e.g., no diving area), or larger-scale management policy area (e.g., regional or national level fishing policies). Individual sites also share similar geographical, environmental, and social features that are possibly dependent on each other. Therefore, estimating the causal impacts of policies such as MPAs requires appropriate methods for clustered and confounded data.

\subsection{Previous Work: Causal Inference in Observational Studies}
Although randomized experiments serve as the gold standard, observational studies can estimate causal effects when all confounding variables are well balanced between treatment groups. To adjust for the imbalance in observed confounding covariates, matching \citep{Stuart:2010} is often applied to isolate causal effects due to its transparency and intuitive appeal. 

While statistical methods to estimate causal effect in observational studies are growing, most methods apply to unstructured data (i.e., without clustering). However, clustering often exists because subjects may be grouped by experimental design, geography, or by sharing higher-level features. Examples include health and educational studies, where patients are nested in hospitals and students are clustered in classrooms or schools. Such clustered data structure poses additional challenges when inferring the causal effect. In our motivating example, the MPA database is naturally clustered, where several sites are nested in the MPA. Capturing MPA-level as well as site-level features (e.g., local social or environmental conditions) is important to remove confounding biases when evaluating the effectiveness of environmental policies. 



To estimate causal effects in clustered data, \citet{Cafri:2019} showed that treatment effect estimation is more accurate when accounting for cluster-level confounding variables. Even if sufficient individual-level covariates are included, ignoring cluster-level confounding covariates would leave a bias on estimation. Within the matching framework, several propensity score methods are developed for the clustered data \citep{Hong:2006, Arpino:2011, Li:2013, Yang:2018}. However, \citet{King:2019} discussed the inefficiency and failure of balancing covariate distributions between treatment groups using the propensity score. They attribute the inefficiency of matching on propensity scores to its goal of mimicking a completely randomized trial rather than a block-randomized trial as well as error in estimating the propensity score.

\subsection{Our Contribution: A Matching Strategy in Clustered Observational Studies}
This article focuses on matching as a nonparametric approach and intuitively mimics a cluster-randomized experiment. We aim to estimate the causal effect by matching estimators under the framework in \citet{Abadie:2006}. Following the characteristics in the MPA database, we analyze the clustered data where the treatment is clustered within the MPA. Nearby sites tend to be assigned the same MPA policy, and one MPA usually contains a single policy only. Cluster-level and unit-level covariates are available, and the outcome is collected at the unit level. To account for the conditional bias when matching on multiple covariates, we adopt the bias-corrected matching estimator \citep{Abadie:2011} in clustered data for two common estimands, the average treatment effect and average treatment effect on the treated, and establish the large sample properties. Under this data structure, matching on cluster-level covariates is sufficient to remove the confounding biases. However, we recommend including relevant unit-level covariates in matching to achieve higher efficiency and lower variance. We show reduced variance in theory and simulation to demonstrate the advantages of matching on both cluster-level and unit-level covariates. 

To account for clustered dependence, we propose a cluster-weighted bootstrap method for variance estimation, which combines the idea of cluster bootstrap \citep{Davison:1997} and weighted bootstrap \citep{Otsu:2017}. Based on a linearization of the matching estimator, the weighted bootstrap method creates residuals so that matching estimators can be viewed as the sample averages of residuals. The variance of the matching estimator can then be approximated by bootstrapping the residuals with appropriate weights. This method preserves the distribution of the number of times that each unit is matched in the resampling procedure. Thus, it avoids the failure of the standard bootstrap in this setting, as discussed in \citet{Abadie:2008}.

The rest of this paper is organized as follows. In Section \ref{s:exploratory}, we introduce the motivating data and describe challenges in establishing causal effects due to the nature of the data structure. Section \ref{s:notation} introduces the notation, assumptions, and estimands of interests. Section \ref{s:ate} explores the large sample properties of matching estimators in clustered data. Section \ref{s:att} presents the cluster-weighted bootstrap procedure for variance estimation. An extension to unit-level treatment assignments for matching estimators is described in Section  \ref{s:extend}. In Section \ref{s:real}, we apply the proposed matching estimator in the MPA data to investigate the causal effect of different marine protection policies on fish biodiversity.  In Section \ref{s:sim}, a simulation study is reported to evaluate the performance of the proposed matching estimator in clustered data. Finally, we conclude our findings in Section \ref{s:disscuss}.

\section{MPA Data and Exploratory Analysis} \label{s:exploratory}
\begin{figure}[H]
\centering
\includegraphics[scale=0.5]{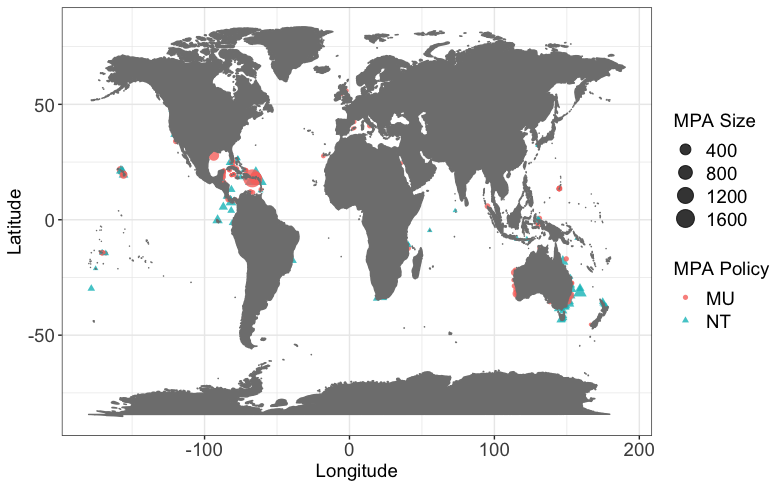}
\caption{Map showing MPA location, size and policy type (MU = multi-use, NT = no-take); MPA policies are present by the majority within each MPA.}
\label{f:map}
\end{figure}

The MPA dataset created by \citet{Gill:2017} includes social, environmental, and ecological information in 9987 sites within 215 MPAs worldwide (Figure \ref{f:map}). The number of sites in each MPA ranges from 1 to 1619, with a mean of 46 and a median of 8. The multi-use (MU) and the no-take (NT) policy  represent two broad categories of types of MPAs. The MU policy  regulates fishing activities to reduce negative impacts,  while the NT policy is more rigorous and prohibits all fishing within the MPA boundaries.  Among 9987 sites, 3988 sites receive the NT policy, whereas 5999 are under the MU policy. The outcome variable is total fish biomass at each site, recorded in underwater visual surveys. There are 13 continuous covariates and 4 categorical covariates that describe the MPA-level and site-level features (Table \ref{t:feature_lsit}). MPA-level covariates include MPA size and country. The other covariates include site-level social and environmental conditions, as well as sampling protocol, location, and date.

\begin{table}[H]
\small\small
\caption{Feature list in the MPA database with units in parentheses for continuous variables and number of levels in parentheses for categorical variables. A detailed summary of the covariates is given in \citet{Gill:2017} Supplementary Table 5.}
\label{t:feature_lsit}
\centering
\begin{tabular}{l|l|l}
\toprule
 & Site-level Covariates & MPA-level Covariates\\
 \midrule
Continuous (13) & \begin{tabular}[c]{@{}l@{}}Latitude (degree)\\Longitude (degree)\\Depth (m)\\ Wave exposure (kW/m)\\Distance to shoreline (km) \\ Distance to population center (``market"; km)\\Coastal population (million/100$\text{km}^2$)\\ Sample date (year) \\ Minimum sea surface temperature ($^{\circ}\text{C}$)\\ Chlorophyll-A ($\text{mg/m}^3$)\\Reef area within 15km ($\text{km}^2$)\\ MPA age (years)\end{tabular} & \begin{tabular}[c]{@{}l@{}} MPA size ($\text{km}^2$)     \end{tabular}             \\
\midrule
Categorical (4) & \begin{tabular}[c]{@{}l@{}}  Habitat type (16) \\ Marine ecoregion (56)\\Sampling protocol (6) \end{tabular} & \begin{tabular}[c]{@{}l@{}} Country (43)\end{tabular}  \\
\bottomrule
\end{tabular}
\end{table}

Sites within the same MPA usually receive the same policy (i.e., same fishing regulations), and each site belongs only to one MPA. As a result, the dataset is naturally clustered where observed sites are nested within the MPA, and conservation policies are geographically clustered. The cluster structure brings difficulty in causal inference due to potential confounding. Both cluster-level and site-level covariates could contribute to the confounding bias. Sites in the same MPA share common environmental, MPA-level and geographical characteristics, affecting both the fish population and MPA policy assignment \citep{Ahmadia:2015, Gill:2017, Ferraro:2019}. Site-specific covariates, including depth, distance to population centers (also called “markets”), size of neighboring human population, and chlorophyll concentration, are also relevant to the ecological outcome \citep{Brewer:2013, Edgar:2014, Gill:2017, Cambell:2020}. Within the same MPA, implementing either the MU or NT policy could be heavily influenced by pre-existing ecological conditions, local tourism, fishing, or politics \citep{Toth:2014, Karr:2015}, which are ideally captured as site-specific covariates. 

\begin{figure}[H]
\centering
\includegraphics[scale=0.5]{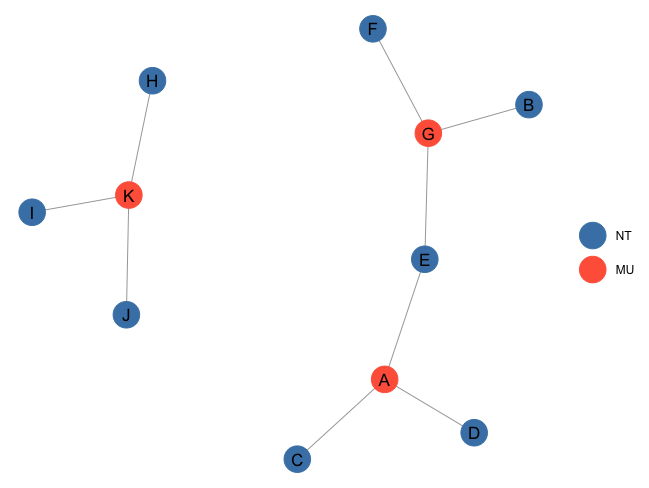}
\caption{A hypothetical example of 1:3 matching of no-take (NT) and multi-use (MT) sites.}
\label{f:match_example}
\end{figure}

Confounding and clustering present two major challenges. We compare the covariate distributions under the two MPA policies for both unadjusted and adjusted samples. The unadjusted sample refers to the raw observation, while the adjusted sample results from multiple matching (one-to-three) using the Mahalanobis distance and with replacement. Figure \ref{f:match_example} is a hypothetical example to illustrate the applied multiple matching. The letters A, G, and K represent sites under the multi-use policy, while the rest of the letters represent sites under the no-take policy. With 1:3 matching, one multi-use site is matched with three no-take sites. Meanwhile, the matched no-take sites can be paired with other multi-use sites. For example,the matched no-take site E is used twice to match site A and G. 

\begin{figure}[H]
\centering
\begin{subfigure}{0.45\textwidth}
\centering
\includegraphics[width=\textwidth]{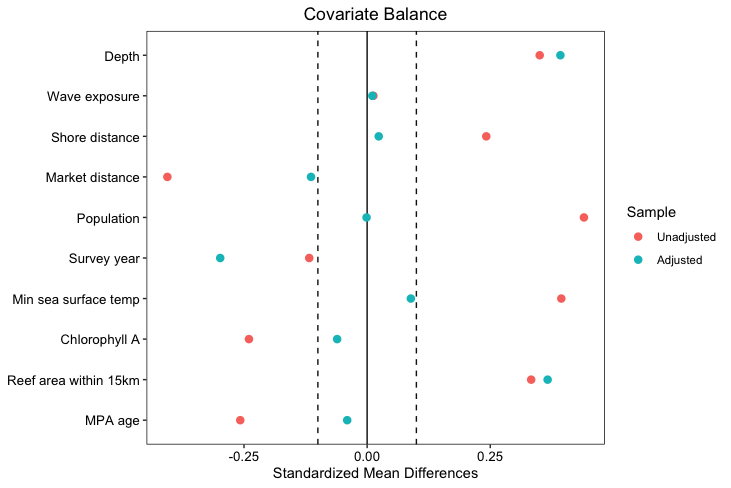}
\caption{Covariate balance between unadjusted and adjusted situations.}
\label{f:covariate_bal}
\end{subfigure}
\begin{subfigure}{0.45\textwidth}
\centering
\includegraphics[width=\textwidth]{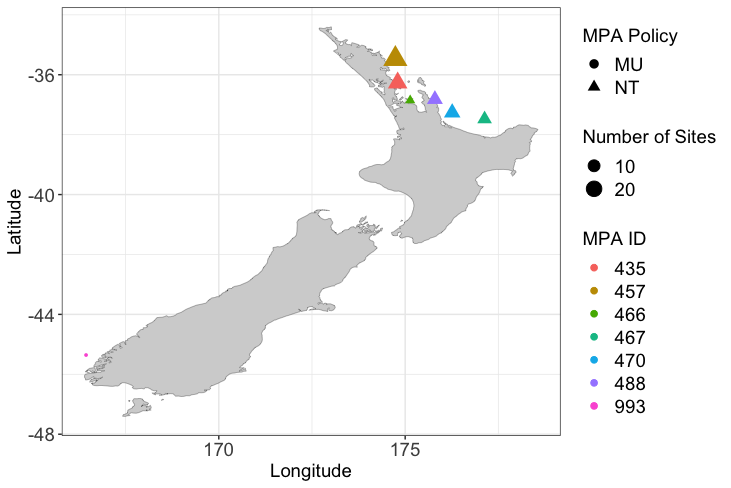}
\caption{MPAs in New Zealand.}
\label{f:nz_cluster}
\end{subfigure}
\caption{Challenges in MPA global dataset.}
\label{f:explore}
\end{figure}

Figure \ref{f:covariate_bal} describes the covariate balance by calculating the standardized mean difference for unadjusted and adjusted samples. For the unadjusted sample, differences between two MPA policies among covariates suggest a  nontrivial impact of confounding. After matching, many covariates become more balanced; however, due to the matching discrepancy (since matching a large dimension of covariates), several covariates still exhibit severe imbalance, requiring further adjustments for residual confounding bias. An example of 7 selected MPA locations in New Zealand is plotted in Figure \ref{f:nz_cluster}, where six are under the NT policy while the other is under the MU policy. Here different MPAs contain different numbers of sites, ranging from 1 to 29. Figure \ref{f:nz_cluster} shows a type of clustering pattern in this MPA global dataset that nearby MPAs tend to follow the same policy. It is practically reasonable because similar regions are likely to share common environmental, geographical and local governmental characteristics, which impact fish biodiversity and MPA policies' choice. Therefore, to explore the causal effect of MPA policies on fish biodiversity, methods that account for both the cluster-level and site-level confounding factors are desired. 

\section{Notation, assumptions and estimands} \label{s:notation}

To establish causal effect in clustered observational studies, we build our proposed matching strategy based on the potential outcomes framework \citep{Holland:1986, Rosenbaum:1983}, also known as the Rubin causal model. Under the potential outcomes framework, the unit-level causal effect is defined as the difference between the outcomes under treatment and control in the same unit. Since we always observe only one of the outcomes for a single unit in reality, it is also considered a missing data problem. Assumptions are needed to estimate the missing counterfactuals. 

The following section briefly introduces the potential outcomes framework for unstructured data and the required assumptions to derive valid causal estimands. Then we extend the framework to clustered data and discuss the corresponding assumptions. 

\subsection{Unstructured Data}
Suppose for unit $i=1,\ldots, N$, $X_i$ is a vector that contains $k$ observed covariates, $A_i\in\{0,1\}$ is the binary treatment indicator and $Y_i$ is the observed outcome. Under the potential outcomes framework, let $Y_i(1)$ be the potential outcome for unit $i$ receiving treatment and $Y_i(0)$ be the potential outcome for unit $i$ receiving control, the observed outcome is $Y_i = Y_i(A_i)$.
This notation implicitly makes the SUTVA \citep{Rubin:1980} that the potential outcomes for each unit are not affected by the treatment assignment of others and there is only one version of the treatment. There are two key components under the SUTVA: (i) no interference and (ii) no alternative versions of the assigned treatment. 

To define the potential outcomes model, we introduce additional notation. 
Given $X_i \in \mathbb{X}$ and $a \in \{0,1\}$, we denote $\mu(x, a)= \mathbb{E}[Y_i|X_i=x, A_i=a], \mu_a(x)=\mathbb{E}[Y_i(a)|X_i=x], \sigma^2(x,a) = \mathbb{V}[Y_i|X_i=x, A_i=a]$ and $\sigma^2_a(x) = \mathbb{V}_a[Y_i(a)|X_i=x]$. We assume the following relationship between the potential outcome $Y_i(a)$ and observed covariates $X_i$, 
\begin{equation} \label{e:model_unstructure}
    Y_i(a) = \mu_{a}(X_i) + \epsilon_i,
\end{equation}
where $\epsilon_i$ independently follows an unspecified distribution with mean 0 and variance $\sigma^2(X_i, a)$ for $i=1,\ldots, N$. Model (\ref{e:model_unstructure}) is flexible because only assumptions for the error term $\epsilon_i$ are imposed. Our goal is to estimate two aggregate estimands, i.e., the average treatment effect and the average treatment effect for the treated. The average treatment effect (ATE) is defined as
$\tau = \mathbb{E}\{Y(1) - Y(0)\}$,
and the average treatment effect on the treated (ATT) is
$\tau^t = \mathbb{E}\{Y(1) - Y(0)|A=1\}$.
We make the following assumption for valid estimation on these two estimands.
\begin{assumption} \label{a:assump-ate}
(i) $\{Y(0),Y(1)\} \perp \!\!\! \perp A|X$; (ii) $\eta < P(A=1|X=x)<1-\eta$ almost surely, for some $\eta>0$.
\end{assumption}

Assumption \ref{a:assump-ate} is often referred to strong ignorability \citep{Rosenbaum:1983}. The first part implies unconfoundness, which is hard to test in nonrandomized studies. It holds that when all confounders are included and controlled as $X$, then conditional on $X$, the treatment assignment is independent of the potential outcomes. The second part states that each unit has a positive probability of receiving treatment or control, which guarantees sufficient overlap in the covariate distribution between groups. Assumption \ref{a:assump-ate} is a strong but fundamental condition to make the ATE and ATT identifiable. Under the assumption of strong ignorability, $\mu(x, a) = \mu_a(x)$ and $\sigma^2(x,a)=\sigma^2_a(x)$. The ATE is then estimated by
$$
\tau = \mathbb{E}[\tau(X)],
$$
where $\tau(X)=\mathbb{E}[Y(1) - Y(0)|X] = \mathbb{E}[Y|X, A=1] - \mathbb{E}[Y|X, A=0]$. 


We present the simple matching estimator proposed in \citet{Abadie:2006}, where matching is done with replacement and by the fixed number of matches from the opposite treatment group. Let $M$ be the number of matches, $\mathcal{J}_M(i)$ contains indices of the $M$ matched units for unit $i$, and $K_M(i)$ is the number of times that unit $i$ is matched, i.e., $K_M(i) = \sum^N_{l=1}\mathbb{I}\{i \in \mathcal{J}_M(l)\}$.
At a unit level where $\tau_i = Y_i(1) - Y_i(0)$, the missing counterfactual can be imputed by the average value of matches as
$$
\hat{Y}_i(0) = \left\{\begin{array}{lr}
     Y_i, & \text{if } A_i=0,  \\
     \frac{1}{M}\sum_{j \in \mathcal{J}_M(i)}Y_j, & \text{if } A_i=1, 
\end{array}\right. \quad \text{and} \quad \hat{Y}_i(1) = \left\{\begin{array}{lr}
      \frac{1}{M}\sum_{j \in \mathcal{J}_M(i)}Y_j, & \text{if } A_i=0, \\
     Y_i, & \text{if } A_i=1.
\end{array}\right.
$$
The estimator for the average treatment effect $\tau$ is 
\begin{equation}
    \hat{\tau}_{\text{mat}} = \frac{1}{N}\sum^N_{i=1}\left\{\hat{Y}_i(1) - \hat{Y}_i(0)\right\} = \frac{1}{N}\sum^N_{i=1}(2A_i - 1)\left\{1+\frac{K_M(i)}{M}\right\}Y_i.
\end{equation}
In \citet{Abadie:2006}, the estimator $\hat{\tau}_{\text{mat}}$ is decomposed into three parts, i.e.,
\begin{equation} \label{e:ate_decompose}
    \hat{\tau}_{\text{mat}} - \tau = \left\{\overline{\tau(X)} - \tau\right\} + E_M + B_M,
\end{equation}
where $\overline{\tau(X)}= \frac{1}{N}\sum^N_{i=1}\left\{\mu_1(X_i) - \mu_0(X_i)\right\}$ represents the average conditional treatment effect,
$E_M =  \frac{1}{N}\sum^N_{i=1}(2A_i - 1)\left\{1+\frac{K_M(i)}{M}\right\}\left\{Y_i - \mu_{A_i}(X_i)\right\}$ is a weighted average of the residuals, 
and $B_M = \frac{1}{N}\sum^N_{i=1}(2A_i - 1)\left[\frac{1}{M}\sum_{j\in \mathcal{J}_M(i)}\left\{\mu_{1 - A_i}(X_i) - \mu_{1-A_i}(X_j)\right\}\right]$ is the conditional bias relative to $\overline{\tau(X)}$.
Based on this decomposition, we have $\mathbb{E}[\overline{\tau(X)} - \tau] = 0$ and $\mathbb{E}[E_M]=0$. The matching discrepancy comes from the conditional bias $B_M$. 

Other than the average treatment effect, researchers are interested in quantifying the causal effect in treated subjects, which is the average treatment effect on the treated (ATT). 
To estimate the ATT, the strong ignorability assumption can be relaxed as follows:
\begin{assumption}\label{a:assump_att}
(i) $Y(0) \perp \!\!\! \perp A|X$; (ii) $P(A=1|X)<1-\eta$ for some $\eta>0$ almost surely.
\end{assumption}
Under Assumption \ref{a:assump_att}, the ATT is $\tau^t = \mathbb{E}[\tau(X)|A=1]$, and its estimator is
\begin{equation}
    \hat{\tau}_{\text{mat}}^t = \frac{1}{N_1}\sum_{A_i=1}\left\{Y_i - \hat{Y}_i(0)\right\} = \frac{1}{N_1}\sum^N_{i=1}\left\{A_i - (1-A_i)\frac{K_M(i)}{M}\right\}Y_i,
\end{equation}
where $N_1$ counts the number of units receiving treatment and $N_0$ is the number of units in the control group. Similar to the decomposition for the ATE estimator in (\ref{e:ate_decompose}),
\begin{equation}
    \hat{\tau}_{\text{mat}}^t - \tau^t = \left\{\overline{\tau(X)}^t - \tau^t\right\} + E_M^t + B_M^t,
\end{equation}
where 
\begin{align*}
    \overline{\tau(X)}^t &= \frac{1}{N_1}\sum^N_{i=1}A_i\left\{\mu(X_i,1) - \mu_0(X_i)\right\},\\
    E_M^t &= \frac{1}{N_1}\sum^N_{i=1}\left\{A_i - (1-A_i)\frac{K_M(i)}{M}\right\}\left\{Y_i - \mu_{A_i}(X_i)\right\},\\
    B_M^t &= \frac{1}{N_1}\sum^N_{i=1}A_i\frac{1}{M}\sum_{j\in \mathcal{J}_M(i)}\left\{\mu_0(X_i) - \mu_0(X_j)\right\}.
\end{align*}

\citet{Abadie:2006} further discusses the conditional biases for the ATE and ATT. For ATE, they show that the order of $\mathbb{E}[B_M]$ is in general not lower than $N^{-2/k}$. The conditional bias could dominate the estimator for large $k$, so that this simple matching estimator is not $N^{1/2}$-consistent. As for the ATT, the conditional bias can be ignored when $N_0$ is in a relatively high order to the number of units being treated $N_1$.

\subsection{Clustered Data}
After reviewing the matching estimator in \citet{Abadie:2006} for independent and identically distributed data, we consider the matching estimator in clustered data. Several works under the potential outcomes framework have been proposed for clustered data \citep{Hong:2006, Stuart:2007, VanderWeele:2008, Li:2013, Yang:2018}. Distinguished from the existing literature,  we focus on the setting where treatment is assigned at the cluster level and leverage the framework of matching estimators in \citet{Abadie:2006} for causal inference. 


For unit $i=1,\ldots,N$ in cluster $r=1,\ldots, R$, we denote $X_{ir}$ as the vector of unit-level covariates, $Z_r$ as the vector of cluster-level covariates, $A_{r}$ as the binary treatment indicator and $Y_{ir}$ as the observed outcome. Let $Y_{ir}(a)$ be the potential outcome for the unit $i$ in the $r$th cluster receiving the treatment $a\in\{0,1\}$, the observed outcome $Y_{ir} = Y_{ir}(A_{r})$. We suppose each unit belongs to only one cluster, clusters are not overlapped, and treatment is assigned at the cluster level, which implies that units within the same cluster are in the same treatment group; see Section \ref{s:extend} for an extension to the case where the treatment varies within clusters. For $X_{ir} \in \mathbb{X}, Z_r \in \mathbb{Z}$ and $a \in \{0,1\}$, we denote $\mu(x, z, a)= \mathbb{E}[Y_{ir}|X_{ir}=x, Z_r=z, A_r=a], \mu_a(x,z)=\mathbb{E}[Y_{ir}(a)|X_{ir}=x, Z_r=z], \sigma^2(x,z, a) = \mathbb{V}[Y_{ir}|X_{ir}=x, Z_r=z, A_{r}=a]$ and $\sigma^2_a(x,z) = \mathbb{V}_a[Y_{ir}(a)|X_{ir}=x,Z_r=z]$.

Similar to the model for unstructured data in (\ref{e:model_unstructure}), we assume the potential outcomes model
\begin{equation} \label{e:model_cluster}
    Y_{ir}(a) = \mu_a(X_{ir}, Z_r) + \alpha_r + \epsilon_{ir},
\end{equation}
where $\alpha_r$ represents unobserved cluster-level random effect with mean 0, $\epsilon_{ir}$ independently follows a distribution with mean 0 and variance $\sigma^2(X_{ir}, Z_r, a)$ for $i=1,\ldots,N$ and $r=1,\ldots, R$. By including the cluster-level random effects, we impose a dependence structure for the error terms. 

We adopt the matching notation for unstructured data and denote $\mathcal{J}_M(i,r)$ as the indices of the $M$ matched units for unit $i$ in the cluster $r$, and $K_M(i,r)$ be the number of times that unit $i$ in cluster $r$ is matched, i.e., $K_M(i,r) = \sum^N_{l=1}\sum^R_{k=1}\mathbb{I}\{({i,r}) \in \mathcal{J}_M(l,k)\}$. Then the missing outcome for the unit $i$ in cluster $r$ can be imputed as 
\[
\hat{Y}_{ir}(0) = \left\{\begin{array}{lr}
     Y_{ir}, & \text{if } A_{r}=0,  \\
     \frac{1}{M}\sum\limits_{(j,k) \in \mathcal{J}_M(i,r)}Y_{jk}, & \text{if } A_{r}=1, 
\end{array}\right. \quad \text{and} \quad 
\hat{Y}_{ir}(1) = \left\{\begin{array}{lr}
      \frac{1}{M}\sum\limits_{(j,k)\in \mathcal{J}_M(i,r)}Y_{jk}, & \text{if } A_{r}=0, \\
     Y_{ir}, & \text{if } A_{r}=1.
\end{array}\right.
\]

To establish a valid causal effect in clustered data, we revisit the necessary assumptions in the unstructured data case. We retain the SUTVA. That is, the potential outcomes for each unit are not influenced by the treatment assigned to other units.  
We then consider the strong ignorability assumption in our setting. Considering that the treatments are assigned at the cluster level, we modify the strong ignorability assumption in unstructured data to accommodate our situation. Under the SUTVA and modified strong ignorability assumptions, $\mu(x, z, a) = \mu_a(x,z)$ and $\sigma^2(x,z,a)=\sigma^2_a(x,z)$. 
\begin{assumption} \label{a:assump_ate_c}
(i) $\{Y_{ir}(0),Y_{ir}(1)\} \perp \!\!\! \perp A_{r}|Z_r$; (ii) $\eta < P(A_{r}=1|Z_r = z)<1-\eta$ almost surely, for some $\eta>0$.
\end{assumption}
Assumption \ref{a:assump_ate_c} suggests conditional independence between treatment $A_{r}$ and the potential outcome $Y_{ir}(a)$ when accounting for the cluster-level covariates $Z_r$. It also requires overlap between treatment groups in the observed covariate distributions. The assumption of strong ignorability in (i) is untestable in practice. However, it is reasonable to proceed under this condition with sufficient cluster-level covariates in the analysis.

The matching estimator for the ATE in clustered data is
\begin{equation}
     \hat{\tau}_{\text{mat}} = \frac{1}{N}\sum^N_{i=1}\sum^R_{r=1}(2A_{r} - 1)\left\{1+\frac{K_M(i,r)}{M}\right\}Y_{ir}.
\end{equation}
Following the similar routine in unstructured data, we decompose the estimator $\hat{\tau}_{\text{mat}}$ as 
\begin{equation}
    \hat{\tau}_{\text{mat}} = \frac{1}{N}\sum^N_{i=1}\sum^R_{r=1}(2A_{r} - 1)\left\{1+\frac{K_M(i,r)}{M}\right\}Y_{ir} = \overline{\tau(X,Z)} + E_M + B_M,
\end{equation}
where 
$$\overline{\tau(X,Z)} = \frac{1}{N}\sum^N_{i=1}\sum^R_{r=1}\left\{\mu_1(X_{ir}, Z_r) - \mu_0(X_{ir}, Z_r)\right\},
$$
$$
E_M = \frac{1}{N}\sum^N_{i=1}\sum^R_{r=1}(2A_{r} - 1)\left\{1+\frac{K_M(i,r)}{M}\right\}\left\{Y_{ir}-\mu_{A_{r}}(X_{ir}, Z_r)\right\},
$$
$$B_M = \frac{1}{N}\sum^N_{i=1}\sum^R_{r=1}(2A_{r} - 1) \left[\frac{1}{M}\sum_{(j,k)\in \mathcal{J}_M(i,r)}\left\{\mu_{1 - A_{r}}(X_{ir},Z_r) - \mu_{1-A_{r}}(X_{jk}, Z_k)\right\}\right].
$$
Consistent with the findings in unstructured data, the matching discrepancy in clustered data comes from $B_M$, which needs careful consideration when matching on multiple covariates.

For the ATT in clustered data, the strong ignorability assumption can be relaxed as in the unstructured data as well.
\begin{assumption} \label{a:assump_att_c}
(i) $\{Y_{ir}(0)\} \perp \!\!\! \perp A_{r}|Z_r$; (ii) $P(A_{r}=1|Z_r = z)<1-\eta$ almost surely, for some $\eta>0$.
\end{assumption}
The estimator of the ATT can be written as,
\begin{equation}
    \hat{\tau}_{\text{mat}}^t = \frac{1}{N_1}\sum_{A_{r}=1}\left\{Y_{ir} - \hat{Y}_{ir}(0)\right\} = \frac{1}{N_1}\sum^N_{i=1}\sum^R_{r=1}\left\{A_{r} - (1-A_{r})\frac{K_M(i,r)}{M}\right\}Y_{ir}.
\end{equation}
Similar to the decomposition for the ATE estimator,
\begin{equation}
    \hat{\tau}_{\text{mat}}^t - \tau^t = \left\{\overline{\tau(X)}^t - \tau^t\right\} + E_M^t + B_M^t,
\end{equation}
where 
\begin{align*}
    \overline{\tau(X)}^t &= \frac{1}{N_1}\sum^N_{i=1}\sum^R_{r=1}A_{r}\left\{\mu(X_{ir},Z_r, 1) - \mu_0(X_{ir}, Z_r)\right\},\\
    E_M^t &= \frac{1}{N_1}\sum^N_{i=1}\sum^R_{r=1}\left\{A_{r} - (1-A_{r})\frac{K_M(i,r)}{M}\right\}\left\{Y_{ir} - \mu_{A_{r}}(X_{ir}, Z_r)\right\},\\
    B_M^t &= \frac{1}{N_1}\sum^N_{i=1}\sum^R_{r=1}A_{r}\frac{1}{M}\sum_{(j,k)\in \mathcal{J}_M(i,r)}\left\{\mu_0(X_{ir}, Z_r) - \mu_0(X_{jk}, Z_{k})\right\}.
\end{align*}

For the ATE and ATT estimators, the matching discrepancy comes from the conditional bias terms $B_M$ and $B_M^t$. According to \citet{Abadie:2006}, $B_M = O_p(N^{-1/k})$ such that the asymptotic distribution of ATE estimator can be dominated by the bias when matching on $k>1$ covariates. Similar rates apply for the ATT estimator, where $B_M^t = O_p(N_1^{-r/k})$ for some $r \geq 1$. Since we consider matching on both cluster-level and unit-level covariates, such bias terms are non-negligible. For instance, in Figure \ref{f:covariate_bal}, although matching reduces covariate imbalance to a large extent, certain covariate imbalance still persists. 
Following \citet{Abadie:2011}, we work on the bias-corrected estimators instead. Denote $\hat{\tau}$ and $\hat{\tau}^t$ as the bias-corrected estiamtors for ATE and ATT, we have
\begin{align*}
   \hat{\tau} &= \hat{\tau}_{\text{mat}} - \hat{B}_M
    \\
    &= \hat{\tau}_{\text{mat}} - \frac{1}{N}\sum^N_{i=1}\sum^R_{r=1}(2A_{r} - 1) \left[\frac{1}{M}\sum_{(j,k)\in \mathcal{J}_M(i,r)}\left\{\hat{\mu}_{1 - A_{r}}(X_{ir},Z_r) - \hat{\mu}_{1-A_{r}}(X_{jk}, Z_k)\right\}\right],
\end{align*}
and
\begin{align*}
    \hat{\tau}^t = \hat{\tau}_{\text{mat}}^t - \hat{B}_M^t = 
     \hat{\tau}_{\text{mat}}^t - \frac{1}{N_1}\sum^N_{i=1}\sum^R_{r=1}A_{r}\frac{1}{M}\sum_{(j,k)\in \mathcal{J}_M(i,r)}\left\{\hat{\mu}_0(X_{ir}, Z_r) - \hat{\mu}_0(X_{jk}, Z_{k})\right\},
\end{align*}
where $\hat{\mu}_{A_{r}}(X_{ir}, Z_r)$ is a consistent estimator of $\mu_{A_r}(X_{ir}, Z_r)$. 

\section{Large Sample Properties} \label{s:ate}
In our clustered data setting, matching on cluster-level covariates is sufficient to account for the confounding variables. However, motivated by blocked randomized experimental designs, matching on both cluster-level and unit-level covariates may improve efficiency. To compare the two matching schemes, we denote $S$ as a unified variable representing either the cluster-level covariates or both cluster-level and unit-level covariates. We explore the asymptotic normality under the fixed $M$ matches for the bias-corrected matching estimators. We then illustrate the benefits of matching on both cluster-level and unit-level covariates based on the asymptotic variance. 

The asymptotic normality under independent and identically distributed data is detailed in \citet{Abadie:2006}. Following \citet{Abadie:2006}, we extend necessary assumptions to clustered data in Assumptions \ref{a:data}. 

\begin{assumption}\label{a:data}
(Conditions for $\hat{\tau}$) 
\begin{enumerate}[label=(\roman*)]
    \item $\{Y_{ir}\}^{N,R}_{i=1,r=1}$ follows Model (\ref{e:model_cluster}) that are independent between clusters but dependent within cluster.
    \item $S$ is continuously distributed on a compact and convex support $\mathbb{S}$. The density of $S$ is bounded and bounded away from zero on $\mathbb{S}$.
    \item $A$ is independent of $(Y(0),Y(1))$ conditional on $S=s$ for almost every $s$. There exists a positive constant $c$ such that $Pr(A=1|S=s) \in (c,1-c)$ for almost every $s$.
    \item For each $a \in \{0,1\}$, $\mu(a,s)$ and $\sigma^2(a,s)$ are Lipschitz in $\mathbb{S}$, $\sigma^2(a,s)$ is bounded away from zero on $\mathbb{S}$ and $E[Y^4|A=a, S=s]$ is bounded uniformly on $\mathbb{S}$.
    \item $E(K_M(i,j)^q)$ is uniformly bounded over $N$. 
\end{enumerate}
\end{assumption}

The bias-corrected estimators involve the estimation of conditional bias terms $B_M$ and $B_M^t$. Following \citet{Abadie:2011}, we assume the required condition for estimation on outcome mean functions in clustered data. 
\begin{assumption}\label{a:mu_series}
(Conditions for $\mu(s,a)$) Let $\lambda = (\lambda_1,\ldots, \lambda_k)'$ be a $k-$dimensional vector of non-negative integers, $\partial^\lambda a(s) = \partial ^{\sum_{l=1}^k\lambda_l}a(s)/\partial s_1^{\lambda_1}\ldots \partial s_k^{\lambda_k}$ and \\$\lvert a(\cdot)\rvert _m = \mbox{max}_{\sum^k_{l=1}\lambda_l \leq m}\mbox{sup}_{s\in \mathbb{S}}\lvert \partial ^\lambda a(s))\rvert$. For each $a\in \{0,1\}$ and $\lambda$ satisfying $\sum^k_{l=1}\lambda_l=k$, the derivative $\partial^\lambda\mu(s, a)$ exists and satisfies $\mbox{sup}_{s \in \mathbb{X}}\lvert \partial^\lambda(\mu(s,a))\rvert \leq C$ for some $C>0$. Furthermore, $\hat{\mu}(s,a)$ satisfies $\lvert \hat{\mu}(\cdot,a) - \mu(\cdot,a)\rvert_{m-1} = o_p(N^{-1/2 + 1/k})$ for each $a\in\{0,1\}$.
\end{assumption}
This assumption guarantees a fast convergence rate on $\hat{B}_M$ and thus contributes to establishing the remarkable result with certain regularity conditions required in \citet{Abadie:2011},
\begin{equation}
    \sqrt{N}(\hat{B}_M - B_M) \stackrel{p}{\rightarrow} 0.
\end{equation}
Assumption \ref{a:mu_series} is also necessary for the bias-corrected ATT estimator $\hat{\tau}^t$ to ensure $\sqrt{N}(\hat{B}_M^t - B_M^t) \stackrel{p}{\rightarrow} 0$ under certain conditions. 

Denote the variance terms
\begin{align*}
    V^{\tau(S)} &= \mathbb{E}[(\tau(S)-\tau)^2], \quad V^{\tau(S),t} = \mathbb{E}[(\tau(S)^t-\tau^t)^2],\\
    V^E &= \mbox{plim}\left[\frac{1}{N^2}\sum^N_{i=1}\left[\sum^R_{r=1}\left\{1+M^{-1}K_M(i,r)\right\}^2\right]\mathbb{V}[Y_{ir}-\mu_{A_{r}}(S_{ir})]\right],\\
    V^{E,t} &= \mbox{plim}\left[\frac{1}{N_1^2}\sum^N_{i=1}\left[\sum^R_{r=1}\left\{A_{r} - (1-A_{r})M^{-1}K_M(i,r))\right\}^2\right]\mathbb{V}[Y_{ir} - \mu_{A_{r}}(S_{ir})]\right],
\end{align*}
we establish the asymptotic normality for bias-corrected estimators in Theorems \ref{thm:ate} and \ref{thm:att}.  

\begin{theorem}\label{thm:ate}
Suppose that Assumptions \ref{a:assump_ate_c}, \ref{a:data} and \ref{a:mu_series} hold. Suppose that the cluster sample size we have, for $r=1,\ldots,R$, satisfy the condition that $\text{min}_{1\leq r \leq R}n_r \rightarrow \infty$ and $\text{sup}_{1\leq r \leq R}n_r = O(N^{1/2})$. Then
$$
\left\{V^E + V^{\tau(S)}\right\}^{-1/2}\sqrt{N}(\hat{\tau} - \tau) \stackrel{d}{\rightarrow} N(0,1),
$$
where $V^E$ and $V^{\tau(S)}$ are finite.
\end{theorem}

\begin{theorem}\label{thm:att}
Suppose that Assumptions \ref{a:assump_att_c}, \ref{a:data} and \ref{a:mu_series} hold. Suppose that the cluster sample size we have for $r=1,\ldots,R$, satisfy the condition that $\mbox{min}_{1\leq r \leq R}n_r \rightarrow \infty$ and $\mbox{sup}_{1\leq r \leq R}n_r = O(N^{1/2})$. Then
$$
\left\{V^{E,t} + V^{\tau(S),t}\right\}^{-1/2}\sqrt{N_1}(\hat{\tau}^t  - \tau^t) \stackrel{d}{\rightarrow} N(0,1),
$$
where $V^{E,t}$ and $V^{\tau(S),t}$ are finite.
\end{theorem}
Proofs for Theorem \ref{thm:ate} and \ref{thm:att} are presented in Supplementary Material \ref{s:app}. 

The asymptotic variances for the two estimators involve the variances of residual terms, i.e., $V^E$ and $V^{E,t}$. Specifically, we have
\begin{align*}
    V^E &= \mbox{plim}\left[\frac{1}{N^2}\sum^N_{i=1}\left[\sum^R_{r=1}\left\{1+M^{-1}K_M(i,r)\right\}^2\right]\mathbb{V}[Y_{ir}-\mu_{A_{r}}(S_{ir})]\right],\\
    V^{E,t}&=\mbox{plim}\left[\frac{1}{N_1^2}\sum^N_{i=1}\left[\sum^R_{r=1}\left\{A_{r} - (1-A_{r})M^{-1}K_M(i,r))\right\}^2\right]\mathbb{V}[Y_{ir} - \mu_{A_{r}}(S_{ir})]\right].
\end{align*}
Consider two matching schemes, i.e., $S = (X, Z)$ and $S = Z$, we note that for $\mathbb{V}[Y_{ir}-\mu_{A_{r}}(X_{ir}, Z_r)]$ and $\mathbb{V}[Y_{ir} - \mu_{A_{r}}(Z_r)]$, when unit-level covariates provide valuable information to the outcome $Y_{ir}$, matching on both cluster-level and unit-level covariates results in lower residual variance than matching on cluster-level covariates. Despite the increased number of covariates for matching, the bias correction helps remove the conditional bias. Hence, we recommend that matching on both cluster-level and unit-level covariates is more efficient when the unit-level information is available. 

\section{Variance Estimation} \label{s:att}
Variance estimation for matching estimators has been investigated in both parametric and nonparametric ways. In \citet{Abadie:2006}, an analytic form for the large sample variance is proposed with consistency achieved. Although the bootstrap method \citep{Efron:1979} is widely used to calculate standard errors for estimators with complicated forms, \citet{Abadie:2008} showed that the standard bootstrap is invalid for matching estimators. The bootstrap method does not preserve the distribution of $K_M(i)$, which follows a Binomial distribution. To overcome this challenge, statistical tools, including wild bootstrap \citep{Huber2016} and weighted bootstrap methods \citep{Otsu:2017}, are developed to estimate asymptotic variance for matching estimators. The weighted bootstrap method \citep{Otsu:2017} can be used when matching is directly performed on covariates, while the wild bootstrap \citep{Huber2016} is applied based on the estimated propensity score. Both methods apply to unstructured data. As for clustered data, there are two main bootstrap strategies: (i) two-stage bootstrap, which is to resample entire clusters at first and then resample subjects within the selected clusters, and (ii) cluster bootstrap, which resamples entire clusters and includes all subjects from the selected clusters. \citet{Davison:1997} discussed different bootstrap methods and showed that the latter is preferable theoretically in clustered data. 

Due to the complex analytic form of matching estimators in clustered data, we employ resampling methods to estimate the variance of the matching estimator and leverage the general procedure of the weighted bootstrap. To account for the dependence in the linear terms in the weighted bootstrap, we incorporate the cluster bootstrap for variance estimation. We describe the extension of weighted bootstrap in clustered data for the ATE first and then present the procedure for our cluster weighted bootstrap method. A similar procedure for the ATT is summarized in Supplementary Material \ref{s:cluster_boot_att}.

Theorem \ref{thm:ate} states that under certain conditions, the bias-corrected estimator $\hat{\tau}$ is asymptotically normal
\[
\frac{\sqrt{N}(\hat{\tau} - \tau)}{\sigma} \stackrel{d}{\rightarrow} N(0,1),
\]
where $\sigma^2 = \sigma_{1}^2 + \sigma_2^2$. The asymptotic variance $\sigma^2$ consists of two parts, i.e.,
$$
\sigma_{1}^2 = \frac{1}{N}\sum^N_{i=1}\sum^R_{r=1}\{1 + M^{-1}K_M(i,r)\}^2\sigma^2(S_{ir}, A_{r}), \quad \sigma_2^2 = E\left[\{\mu(S_{ir}, 1) - \mu(S_{ir}, 0)-\tau\}^2\right],
$$
which measure the variability of $\overline{\tau(S)} - \tau$ and weighted average of the residuals $E_M$. Following \citet{Otsu:2017}, we write our bias-corrected estimator $\hat{\tau} =\hat{\tau}_{\text{mat}} - \hat{B}_M$ as a linear form in clustered data, 
\begin{align*}
    \hat{\tau} &=\hat{\tau}_{\text{mat}} - \hat{B}_M\\
    &=  \frac{1}{N}\sum^N_{i=1}\sum^R_{r=1}\left\{\hat{\mu}_1(S_{ir}) - \hat{\mu}_0(S_{ir})\right\} +  \frac{1}{N}\sum^N_{i=1}\sum^R_{r=1}(2A_{r} - 1)\left\{1+\frac{1}{M}K_M(i,r)\right\}\left\{Y_{ir} - \hat{\mu}_{A_{r}}(S_{ir})\right\}\\
    &=   \frac{1}{N}\sum^N_{i=1}\sum^R_{r=1}\left[\left\{\hat{\mu}_1(S_{ir}) - \hat{\mu}_0(S_{ir})\right\} + (2A_{r} - 1)\left\{1+\frac{1}{M}K_M(i,r)\right\}\left\{Y_{ir} - \hat{\mu}_{A_{r}}(S_{ir})\right\} \right]\\
    &=   \frac{1}{N}\sum^N_{i=1}\hat{\tau}_i
\end{align*}
where $\hat{\tau}_i = \sum^R_{r=1}\left\{\hat{\mu}_1(S_{ir}) - \hat{\mu}_0(S_{ir})\right\} + (2A_{r} - 1)\left\{1+\frac{1}{M}K_M(i,r)\right\}\left\{Y_{ir} - \hat{\mu}_{A_{r}}(S_{ir})\right\}$. 

Based on this form, we express the $i$th residual as
$$
\hat{\tau}_i - \hat{\tau} = \hat{e}_i + \hat{\xi}_i,
$$
with $\hat{e}_i$ and $\hat{\xi}_i$ are estimated values of $e_i = \sum^R_{r=1}(2A_{r} - 1)\left\{1+\frac{K_M(i,r)}{M}\right\}\{Y_{ir} - \mu_{A_{r}}(S_{ir})\}$ and $\xi_i = \sum^R_{r=1}\{\mu_{A_{r}}(S_{ir}) - \mu_{1-A_{r}}(S_{ir})\} - \tau$, respectively. This form is the key aspect to obtain the estimated variance for matching estimators. 
Therefore, treating $\{\hat{\tau}_i\}^N_{i=1}$ as observations and bootstrapping on $\{\hat{\tau}_i\}^N_{i=1}$ would lead to a valid bootstrap variance estimation. 

Considering the clustered data where unobserved cluster-level confounding covariates could exist across and within clusters, we adopt the cluster bootstrap method \citep{Davison:1997} to account for the variance from potential cluster-level covariates. The cluster bootstrap suggests resampling on the cluster levels first and then including all observations within selected units. Incorporating the weighted bootstrap algorithm, we propose our cluster-weighted bootstrap method as follows.
\begin{itemize}
    \item \textit{\textbf{Step 1}}: Obtain the weighted bootstrap samples $\{\hat{\tau}_i\}^N_{i=1}$ based on the matching estimator framework.
    \item \textit{\textbf{Step 2}}: For clustered data with $n$ observations and $R$ non-overlapped clusters, sample $R$ clusters with replacement.
    \item \textit{\textbf{Step 3}}: Include all $\{\hat{\tau}_i\}^N_{i=1}$ within selected clusters and calculate their corresponding weights $\{W_i^*\}^N_{i=1}$. One option of generating weights is to set $W_i^*=M_i^*/\sqrt{N}$, where $(M_1^*, \ldots, M_N^*)$ is a vector from a multinomial distribution with equal probability. 
    \item \textit{\textbf{Step 4}}: Obtain a bootstrap replicate as $\hat{\tau}^*_b = \sum^N_{i=1}W_i^*(\hat{\tau}_i - \hat{\tau})$.
    \item \textit{\textbf{Step 5}}: Repeat the Step 1-4 $B$ times. Compute the bootstrap variance estimator for the bias-corrected matching estimator $\hat{\tau}$ as the empirical variance of $\{\hat{\tau}^*_b\}^B_{b=1}$.
\end{itemize}

\begin{theorem}\label{thm:asymp_valid}
Suppose that Assumptions \ref{a:assump_ate_c}, \ref{a:data}, and Assumption \ref{a:weight} in Supplementary Material \ref{s:app_asymp} hold. Suppose that the cluster sample size we have, for $r=1,\ldots,R$, satisfy the condition that $\text{min}_{1\leq r \leq R}n_r \rightarrow \infty$ and $\text{sup}_{1\leq r \leq R}n_r = O(n^{1/2})$. Then
$$
E\left[\left\{\sum^N_{i=1}W_i^*(\hat{\tau}_i - \hat{\tau})|(\bm{Y},\bm{A},\bm{S})\right\}^2\right] \stackrel{p}{\rightarrow} \sigma^2
$$
where $\sigma^2$ is the asymptotic variance of cluster matching estimator $\tilde{\tau}$.
\end{theorem}
Proof for Theorem \ref{thm:asymp_valid} is detailed in Supplementary Material \ref{s:app_asymp}.

\section{Extension to unit-level treatment assignments} \label{s:extend}

Although we focus on the case with cluster treatment assignments (i.e., sites within a cluster receive the same treatment), our framework extends readily to the case with unit-level treatment assignments (i.e., sites with a cluster can receive different treatments). In the motivating application, most MPAs have cluster treatment, but some MPAs have a few, although a small number of sites under a different policy from most sites. It is because some MPAs contain multiple zones with different regulations \citep{Barbara:2016}. To accommodate this situation, it is sufficient to make simple modifications for applying the above method, theory, and inference. First, we change the cluster-level treatment $A_r$ to the unit-level treatment $A_{ir}$. The ignorability assumptions for the ATE and ATT become the following. 

\begin{assumption}
(i) $\{Y_{ir}(0),Y_{ir}(1)\} \perp \!\!\! \perp A_{ir}|X_{ir},Z_r$; (ii) $\eta < P(A_{ir}=1|X_{ir}=x, Z_r=z)<1-\eta$ almost surely, for some $\eta>0$.
\end{assumption}

\begin{assumption}
(i) $\{Y_{ir}(0)\} \perp \!\!\! \perp A_{ir}|X_{ir}, Z_r$; (ii) $P(A_{ir}=1|X_{ir}=x, Z_r = z)<1-\eta$ almost surely, for some $\eta>0$.
\end{assumption}

In the matching procedure, matching based on cluster-level confounders is insufficient to remove confounding biases. Thus we use both cluster-level and site-level confounders in matching, which is also recommended for the case with cluster treatment assignments. The role of site-level confounders removes confounding biases instead of improving the efficiency of the matching estimator.


\section{Analysis of Conservation Policy Effects on Marine Biodiversity} \label{s:real}

The proposed matching framework is applied to the \citet{Gill:2017} MPA dataset to investigate the causal relationship between MPA policies and fish biodiversity. We consider the multi-use policy (MU) as the treatment while the no-take policy (NT) serves as the control. Then the following two estimands are of interest, 
\begin{align*}
    \mbox{ATE} &= \mathbb{E}\left[Y(\mbox{MU}) - Y(\mbox{NT})\right],\\
    \mbox{ATT} &= \mathbb{E}\left[Y(\mbox{MU}) - Y(\mbox{NT})|A = \mbox{MU}\right].
\end{align*}
We impute the missing counterfactual for each site by matching with replacement and with a fixed number of matches. The Mahalanobis distance metric is calculated to measure the similarity of sites under different policies. Three nearest control sites are matched to the treatment site. Due to the skewed covariates distributions between two groups, we transform the non-negative continuous covariates by the Box-Cox transformation. The outcome of interest is the log transformation of fish biomass ($\text{g/100m}^2$). Following our recommendation, matching is conducted by both MPA-level and site-level covariates, and the performance is compared to matching by MPA-level covariates. The mean outcome functions are estimated by three methods: the spline regression \citep{Schoenberg1946}, the sieve method \citep{Chen:2007} and the regression forest \citep{Athey:2019}. Because of the large number of first-order and second-order terms among covariates, we apply the LASSO method \citep{Tibshirani:1996} to select a subset of covariates before matching. The regularization parameter is chosen by 5-fold cross validation and the bias-corrected estimators for the ATE and ATT are derived. Corresponding variances are computed by the proposed cluster-weighted bootstrap method. 


Table \ref{t:real_result} summarizes the estimated ATE and ATT with 95\% confidence intervals under different matching strategies. When matching only on MPA-level covariates, all three methods result in negative point estimates for ATE and ATT. However, all three methods detect no significant differences between the MU policy and NT policy on fish biomass. When matching on both MPA-level and site-level covariates, the sieve method shows significantly different impacts of MPA policies on fish biomass. The ATE under the sieve method is significantly negative, suggesting that the NT policy is more beneficial to fish biodiversity. The result of the ATT by using the sieve method is consistent with the ATE result, which shows a positive impact of using the NT policy. Comparing the procedures between matching on MPA-level covariates and matching on all relevant covariates, we find a reduced estimated variance under the sieve method as expected when including individual covariates that are relevant to the outcome. 

In this MPA data, several site-level covariates such as distance to markets, reef area within 15 km, and neighboring human population size are relevant to the fish biodiversity under either the MU or NT policy (See Figure \ref{f:imp_plot} in Supplementary Material \ref{s:sup_figure}). Certain regions at the habitat, country and ecoregion levels also show an important association with the fish biodiversity. Taking these covariates into account for matching helps reduce the matching variance. Using the regression forest method also reveals the beneficial effect of the NT policy for the estimated ATE, while the effect is barely detected for the ATT. 


\begin{table}[H]
\caption{Summary of the average treatment effect (ATE) and the average treatment effect on the treated (ATT) with estimated standard errors in parentheses when comparing the multi-use (MU) policy and no-taken (NT) policy in MPAs where MU is considered as treatment group; Response is $\log{(\mbox{Fish Biomass})}$.}
\label{t:real_result}
\centering
\resizebox{\textwidth}{!}{
\begin{tabular}{r|cccc}
\toprule
\multicolumn{1}{c|}{}                                                & \multicolumn{2}{c}{ATE}        & \multicolumn{2}{c}{ATT}         \\
\multicolumn{1}{c|}{}                                                & Point Estimate & 95\% CI        & Point Estimate & 95\% CI        \\ 
\midrule
\multicolumn{1}{l|}{Matching on MPA-level covariates}                &                &                &                &                \\
Sieve Method                                                         & -0.49 (0.31)   & (-1.10, 0.13)  & -0.67 (0.50)   & (-1.64, 0.30)  \\
Smooth Spline                                                        & -0.27 (0.25)   & (-0.77, 0.23)  & -0.19 (0.38)   & (-0.93, 0.56)  \\
Regression Forest                                                    & -0.57 (0.35)   & (-1.26, 0.12)  & -0.82 (0.49)   & (-1.77, 0.13)  \\
\multicolumn{1}{l|}{Matching on all covariates} &                &                &                &                \\
Sieve Method                                                         & -0.41 (0.17)   & (-0.76, -0.07) & -0.58 (0.26)   & (-1.10, -0.06) \\
Smooth Spline                                                        & -0.34 (0.30)   & (-0.93, 0.26)  & -0.41 (0.41)   & (-1.22, 0.40)  \\
Regression Forest                                                    & -0.70 (0.32)   & (-1.32, -0.07) & -1.03 (0.53)   & (-2.06, 0.00)  \\
\bottomrule
\end{tabular}}
\end{table}


\section{Simulation Study} \label{s:sim}
We conduct a simulation study to evaluate the finite-sample performance of matching estimators when matching on cluster-level covariates only or matching on both cluster-level and unit-level covariates, and compare the proposed cluster weighted bootstrap method with the original weighted bootstrap on variance estimation. 

\subsection{Setting}
Let the number of clusters be $R=50$ and consider two clustered data settings: (i) balanced clustered data, and (ii) unbalanced clustered data. For balanced clustered data, all clusters have the same size and we vary the common cluster size $n_r \in \{10, 50, 100\}$. In the unbalanced case, we randomly choose $n_r$ from a discrete uniform distribution $\text{Unif\{20,100\}}$ independently across clusters. Unit-level confounding variables $X_{ir} \in \mathbb{R}^6$ are generated independently from a uniform distribution $\mbox{Unif}(-1,1)$. We assume observations in the same cluster have the same treatment, and include one cluster-level covariate $Z_r \sim \mbox{Unif}(0,1)$. We assume nonlinear relationships between $X_{ir}, Z_r$ and the potential outcomes. Specifically, we consider the following transformations,
\begin{align*}
    X_1^* &= g(X_1)g(X_2), \quad X_2^* = g(X_1) + g(X_2), \quad X_3^* = 3\mbox{max}(X_3, 0),\\
    X_4^* &= 3\mbox{max}(X_4, 0), \quad X_5^* = 3\mbox{max}(X_5, 0), \quad X_6^* = 2X_6 - 1, Z_r^* = g(Z_r),
\end{align*}
where $g(x) = 1 + \left[1 + \exp{\left\{-20 (x-1/3)\right\}}\right]^{-1}$.
The transformed variables $\{X_j^*\}^6_{j=1}$ and $Z_r^*$ are standardized with mean 0 and variance 1. 

We estimate the ATE and ATT by the proposed matching estimators. To make the situation more complex, a random effect that represents unobserved cluster-level confounding covariates is included with $\alpha_r \stackrel{\text{iid}}{\sim} N(0,1)$ for $r=1,\ldots, R$. Let regression coefficients $\bbeta = (1,1,1,1,1,1)^T$, true treatment effect $\gamma=2$ and random error $\epsilon_{ir} \stackrel{\text{iid}}{\sim} N(0,1)$.
Simulated data are then generated based on the following mechanism for $\{Y_{ir}, A_{r}, X_{ir}, Z_r\}^{n,R}_{i=1,r=1}$,
$$
Y_{ir}(0) = \bm{X}^*\bm{\beta}  + Z_r^* + \alpha_r + \epsilon_{ir}, \quad Y_{ir}(1) =\bm{X}^*\bm{\beta} + Z_r^* + \gamma + \alpha_r + \epsilon_{ir}.
$$
The treatment indicator $A_{r}$ is assigned at cluster level following a Binomial distribution, $\mbox{Bin}\left\{\pi(Z_r)\right\}$, where $\pi(Z_r) = (1 + f(Z_r; 2, 4))/4$ and $f(2,4)$ is the Beta cumulative distribution function. 

Matching is performed through the R package \texttt{Matching}. We set the number of matches to $M=3$ and use the Mahalanobis distance as the measuring metric. The outcome mean functions $\mu_1(X_{ir}, Z_{r})$ and $\mu_0(X_{ir}, Z_{r})$ can be estimated by various ways. Here we consider three methods that are utilized in the real data analysis: (i) the sieve method, (ii) linear spline method, and (iii) generalized regression forest for bias correction. Corresponding estimators are denoted as $\hat{\tau}_{\text{sieve}}, \hat{\tau}_{\text{ls}}, \hat{\tau}_{\text{rf}}$, respectively. 

We compare the estimation performance in the following procedures:
\begin{itemize}
    \item Procedure 1: match on the cluster-level covariate and estimate variance by the proposed cluster-weighted bootstrap method,
    \item Procedure 2: match on the cluster-level covariate and estimate variance by the standard weighted bootstrap method,
    \item Procedure 3: match on both cluster-level and unit-level covariates and estimate variance by the proposed cluster-weighted bootstrap method,
    \item Procedure 4: match on both cluster-level and unit-level covariates and estimate variance by the standard weighted bootstrap method.
\end{itemize}
Bootstrap is performed with $B=1000$ replicates. Performance of matching estimators are evaluated on the average values of biases, variances and coverages for 95\% confidence intervals based on 1000 simulated datasets.

\subsection{Results}
Table \ref{t:ate} summarizes the average values of biases, variances, and coverages of 95\% confidence intervals under different scenarios for the ATE. When matching on the cluster-level covariate, the small biases for both balanced and unbalanced cluster size settings suggest that matching on the cluster-level covariate is sufficient to remove estimation bias for the treatment effect. It is also consistent with the theoretical findings in \citet{Abadie:2006} that the conditional bias is ignorable when matching on a single covariate. When comparing three approaches to approximate the conditional outcome mean functions, the sieve method usually outperforms the other two with the lowest bias. The linear spline method shows advantages in reducing the bias for unbalanced data, but suffers from the underestimated variance. The regression forest method has the largest absolute bias in most cases. 

As for variance estimation, the standard weighted bootstrap always results in a much smaller variance than the cluster-weighted bootstrap and thus lower coverage for the 95\% confidence interval. By ignoring the cluster effect, the weighted bootstrap method resamples on the unit level such that it fails in approximating the true distribution of matching estimators. On the contrary, by taking clusters into consideration and resampling at the cluster level, our proposed cluster-weighted bootstrap method results in high coverage for the 95\% confidence interval. 


Comparing Procedures 1 and 3, the estimated variances when using both cluster and unit covariates for matching are always smaller than those using only the cluster-level covariate for matching. Though matching on the cluster-level covariate leads to small biases, the consistently reduced estimated variances suggest that matching on both cluster-level and unit-level covariates is beneficial and can achieve high coverage for the 95\% confidence interval.


When estimating the ATT (Table \ref{t:att}), the sieve method still achieves the most accurate estimation and usually the highest coverage probability. When matching on both cluster and unit covariates, the estimated variance is lower than matching on the cluster-level covariate, and the coverages for the 95\% confidence interval are as close to 95\% as those when matching on the cluster-level covariate. 

\begin{table}[H]
\centering
\caption{Summary of bias ($\times 10^3$), average variance ($\times 10^3$) and coverage (\%) of 95\% confidence intervals under different number of clusters $R$ and different cluster size $n_r$ when estimating the average treatment effect (ATE) based on 1000 Monte Carlo samples; matching is performed by using cluster-level covariate only or using both cluster-level and unit-level covariates; variance is estimated by the cluster-weighted bootstrap or the standard weighted bootstrap method.}
\label{t:ate}
\begin{tabular}{lcccccccccccccccccc} 
\toprule
$(R, n_r)$                                         & \multicolumn{3}{c}{$(50,10)$} &  & \multicolumn{3}{c}{$(50,50)$} &  & \multicolumn{3}{c}{$(50,100)$} &  & \multicolumn{3}{c}{$(50,[20,100])$}  \\
                                                      & bias   & var   & cvg            &  & bias & var & cvg                &  & bias & var & cvg                &  & bias   & var   & cvg                   \\
                                                      \midrule
\multicolumn{16}{c}{Procedure 1: matching on cluster covariate only \& cluster-weighted bootstrap}                                                                                                                                                                           \\[0.2cm]
$\hat{\tau}_{\text{sieve}}$ & -15 & 215 & 96.1           &  & 17     & 170    & 95.2                  &  & -4     & 158    & 94.7                    &  & -15 & 172 & 95.4                  \\
$\hat{\tau}_{\text{ls}}$    & 33  & 177 & 93.4           &  & 56     & 148    & 90.8                   &  & 28    & 136    & 92.3                     &  & 13  & 145 & 92.8                  \\
$\hat{\tau}_{\text{rf}}$    & -87 & 281 & 97.0           &  & -35     & 143    & 91.8                   &  & -49     & 114    & 89.9                    &  & -69 & 138 & 92.7   \\[0.2cm]
\multicolumn{16}{c}{Procedure 2: matching on cluster covariate only \& standard weighted bootstrap}                                                                                                                                                                           \\[0.2cm]
$\hat{\tau}_{\text{sieve}}$ & -15 & 97 & 85.2           &  & 17     & 21    & 53.2                   &  & -4     & 10    & 40.6                    &  & -15 & 20 & 54.2                  \\
$\hat{\tau}_{\text{ls}}$    & 33  & 66 & 77.5           &  & 56     & 14    & 43.3                   &  & 28     & 7    & 34.4                    &  & 13  & 13 & 45.3                  \\
$\hat{\tau}_{\text{rf}}$    & -87 & 127 & 88.4           &  & -35     & 15    & 48.1                   &  & -49     & 6     & 33.3                     &  & -69 & 14 & 48.1   \\[0.4cm]
\multicolumn{16}{c}{Procedure 3: matching on both cluster and unit covariates \& cluster-weighted bootstrap}                                                                                                                                                           \\[0.2cm]
$\hat{\tau}_{\text{sieve}}$ & -113 & 159 & 94.0           &  & -59     & 119    & 92.6                   &  & -64     & 112    & 92.3                    &  & -76 & 129 & 93.9                  \\
$\hat{\tau}_{\text{ls}}$    & 33  & 125 & 91.8           &  & 65     & 104    & 91.0                   &  & 47     & 100    & 90.2                    &  & 39  & 110 & 91.1                  \\
$\hat{\tau}_{\text{rf}}$    & -243 & 184 & 93.2           &  & -111     & 99    & 89.4                   &  & -91     & 85    & 89.0                    &  & -128 & 103 & 90.3                  \\[0.2cm]
\multicolumn{16}{c}{Procedure 4: matching on both cluster and unit covariates \& standard weighted bootstrap}                                                                                                                                                           \\[0.2cm]
$\hat{\tau}_{\text{sieve}}$ & -113 & 73 & 80.0           &  & -59     & 15    & 50.1                   &  & -64     & 8    & 35.1                    &  & -76 & 13 & 47.6                  \\
$\hat{\tau}_{\text{ls}}$    & 33  & 49 & 73.6           &  & 65     & 10    & 41.4                   &  & 47     & 5    & 29.9                    &  & 39  & 9 & 42.5                  \\
$\hat{\tau}_{\text{rf}}$    & -243 & 87 & 79.4           &  & -111     & 11    & 42.5                   &  & -91     & 5    & 29.1                    &  & -128 & 9 & 40.6                  \\
\bottomrule
\end{tabular}
\end{table}

\begin{table}[H]
\centering
\caption{Summary of bias ($\times 10^3$), average variance ($\times 10^3$) and coverage (\%) of 95\% confidence intervals under different number of clusters $R$ and different cluster size $n_r$ when estimating the average treatment effect on the treated (ATT) based on 1000 Monte Carlo samples; matching is performed by using cluster-level covariate only or using both cluster-level and unit-level covariates; variance is estimated by the cluster-weighted bootstrap or the standard weighted bootstrap method.}
\label{t:att}
\begin{tabular}{lcccccccccccccccccc} 
\toprule
$(R, n_r)$                                         & \multicolumn{3}{c}{$(50,10)$} &  & \multicolumn{3}{c}{$(50,50)$} &  & \multicolumn{3}{c}{$(50,100)$} &  & \multicolumn{3}{c}{$(50,[20,100])$}  \\
                                                      & bias   & var   & cvg            &  & bias & var & cvg                &  & bias & var & cvg                &  & bias   & var   & cvg                   \\
                                                      \midrule
\multicolumn{16}{c}{Procedure 1: matching on cluster covariate only \& cluster-weighted bootstrap}                                                                                                                                                                           \\[0.2cm]
$\hat{\tau}_{\text{sieve}}$ & 50 & 305 & 96.8           &  & 17     & 231    & 96.3                   &  & 11     & 226    & 95.9                     &  & 15 & 248 & 97.0                  \\
$\hat{\tau}_{\text{ls}}$    & 33  & 227 & 93.8           &  & 37     & 189    & 92.7                   &  & 33     & 183    & 91.9                    &  & 30  & 203 & 93.8                  \\
$\hat{\tau}_{\text{rf}}$    & -126 & 358 & 96.6           &  & -101     & 169    & 91.7                   &  & -99     & 141    & 90.3                    &  & -108 & 178 & 92.9   \\[0.2cm]
\multicolumn{16}{c}{Procedure 2: matching on cluster covariate only \& standard weighted bootstrap}                                                                                                                                                                           \\[0.2cm]
$\hat{\tau}_{\text{sieve}}$ & 50 & 138 & 89.4           &  & 17     & 29    & 52.2                  &  & 11     & 15    & 42.3                    &  & 15 & 29 & 53.5                  \\
$\hat{\tau}_{\text{ls}}$    & 33  & 89 & 80.1           &  & 37     & 19    & 42.7                   &  & 33     & 10    & 33.7                    &  & 30  & 19 & 44.8                  \\
$\hat{\tau}_{\text{rf}}$    & -126 & 168 & 89.0           &  & -101     & 20    & 46.5                   &  & -99     & 8    & 32.0                    &  & -108 & 19 & 47.7   \\[0.4cm]
\multicolumn{16}{c}{Procedure 3: matching on both cluster and unit covariates \& cluster-weighted bootstrap}                                                                                                                                                           \\[0.2cm]
$\hat{\tau}_{\text{sieve}}$ & 228 & 195 & 92.8           &  & 108     & 139    & 93.4                   &  & 63     & 131    & 93.9                     &  & 88 & 152 & 95.5                  \\
$\hat{\tau}_{\text{ls}}$    & 46  & 141 & 90.6           &  & 8     & 114    & 90.7                   &  & -7     & 110    & 89.1                    &  & -3  & 122 & 90.0                  \\
$\hat{\tau}_{\text{rf}}$    & -243 & 216 & 97.2           &  & -93     & 107    & 90.8                   &  & -89     & 92    & 88.9                    &  & -104 & 113 & 91.2                  \\[0.2cm]
\multicolumn{16}{c}{Procedure 4: matching on both cluster and unit covariates \& standard weighted bootstrap}                                                                                                                                                           \\[0.2cm]
$\hat{\tau}_{\text{sieve}}$ & 228 & 92 & 79.7           &  & 108     & 18    & 50.9                   &  & 63     & 9    & 39.8                     &  & 88 & 16 & 48.6                  \\
$\hat{\tau}_{\text{ls}}$    & 46  & 58 & 74.5           &  & 8     & 12    & 41.7                   &  & -7     & 6    & 30.1                    &  & -3  & 10 & 41.8                  \\
$\hat{\tau}_{\text{rf}}$    & -243 & 105 & 89.8           &  & -93     & 13    & 44.5                   &  & -89     & 6    & 29.3                    &  & -104 & 10 & 42.3                  \\
\bottomrule
\end{tabular}
\end{table}

\section{Discussion} \label{s:disscuss}
In this paper, we consider matching in a nonparametric way and discuss the matching estimators for causal inference in clustered observational studies. Large sample properties for two estimands of interest are explored. For variance estimation, we propose a cluster-weighted bootstrap method that avoids the failure of the standard bootstrap and adjusts for the cluster effect. When treatment assignment occurs at the cluster level, balancing on cluster-level covariates is sufficient to remove confounding biases. However, we recommend that matching on both cluster-level and unit-level covariates is more efficient. 

We demonstrate the advantages of matching on both cluster-level and unit-level covariates in theory and simulation. Simulation results show reduced variance when matching on both cluster and unit-level covariates in various settings. Compared to the standard weighted bootstrap method widely applied in unstructured data, the proposed cluster-weighted bootstrap outperforms with much higher coverages of 95\% confidence interval. Three methods are utilized in constructing the bias-corrected estimators. In our result, the sieve method frequently achieves higher 95\% confidence interval coverages and lower biases than the others. 

We apply the recommended matching strategy to study the ecological effects of different marine protected area policies on fish biodiversity. When matching on both MPA-level and site-level covariates, the sieve method with cluster-weighted bootstrap successfully detects different ecological impacts between the multi-use and no-take policies. Consistent with the results in \citet{Gill:2017}, we find that the no-take policy positively affects the fish population compared to the multi-use policy. However, there remain several undiscussed aspects. First, the causal effect is estimated under strong assumptions. The assumption of SUTVA may be violated when there are multiple versions of multi-use and no-take policies. Second, as discussed in \citet{Gill:2017}, conservation outcomes are closely related to the MPA management processes. Variability in MPA management effectiveness may cause different conservation impacts for the same MPA policy. In future studies, covariates that measure the adequacy and appropriateness of management should be included when comparing the relative causal effects of different MPA policies.

There are several issues that warrant future research. Our current analysis and recommendation are based on the clustered data with a binary treatment assignment. This method could be generalized to a complicated data structure involving multilevel clusters with more than two treatment groups or continuous treatments. The benefits of matching estimators by different layers of observed covariates remain unknown. Nonparametric methods for variance estimation in multilevel observational studies would be complicated. Our proposed cluster-weighted bootstrap may be extended to these settings. As with most causal inference methods, we require the SUTVA and strong ignorability of treatment assumptions, which may be violated in some circumstances. It is of interest to study matching estimators in clustered data under relaxed assumptions and develop sensitivity analysis \citep{Yang:2018b} to assess the robustness of the study conclusions to key assumptions. Also, in cases with extremely high-dimensional $X$, matching on $X$ even with bias correction may not be effective. One can use the double score matching idea \citep{Yang:2020} to conduct dimension reduction prior to matching. 
\bibliographystyle{rss.bst}
\bibliography{match.bib}

\begin{thebibliography}{39}
\expandafter\ifx\csname natexlab\endcsname\relax\def\natexlab#1{#1}\fi
\expandafter\ifx\csname url\endcsname\relax
  \def\url#1{\texttt{#1}}\fi
\expandafter\ifx\csname urlprefix\endcsname\relax\def\urlprefix{URL}\fi

\bibitem[{Abadie and Imbens(2006)}]{Abadie:2006}
Abadie, A. and Imbens, G.~W. (2006) Large sample properties of matching
  estimators for average treatment effects.
\newblock \textit{Econometrica}, \textbf{74}, 235--267.

\bibitem[{Abadie and Imbens(2008)}]{Abadie:2008}
--- (2008) On the failure of the bootstrap for matching estimators.
\newblock \textit{Econometrica}, \textbf{76}, 1537--1557.

\bibitem[{Abadie and Imbens(2011)}]{Abadie:2011}
--- (2011) Bias-corrected matching estimators for average treatment effects.
\newblock \textit{Journal of Business \& Economic Statistics}, \textbf{29},
  1--11.

\bibitem[{Ahmadia et~al.(2015)Ahmadia, Glew, Provost, Gill, Hidayat, Mangubhai,
  Purwanto and Fox}]{Ahmadia:2015}
Ahmadia, G., Glew, L., Provost, M., Gill, D., Hidayat, N., Mangubhai, S.,
  Purwanto, P. and Fox, H. (2015) Integrating impact evaluation in the design
  and implementation of monitoring marine protected areas.
\newblock \textit{Philosophical Transactions B}, \textbf{370}.

\bibitem[{Arpino and Mealli(2011)}]{Arpino:2011}
Arpino, B. and Mealli, F. (2011) The specification of the propensity score in
  multilevel observational studies.
\newblock \textit{Computational Statistics \& Data Analysis}, \textbf{55},
  1770--1780.

\bibitem[{Athey et~al.(2019)Athey, Tibshirani and Wager}]{Athey:2019}
Athey, S., Tibshirani, J. and Wager, S. (2019) Generalized random forests.
\newblock \textit{The Annals of Statistics}, \textbf{47}, 1148--1178.

\bibitem[{Bennett and Dearden(2014)}]{Bennett:2014}
Bennett, N.~J. and Dearden, P. (2014) Why local people do not support
  conservation: Community perceptions of marine protected area livelihood
  impacts, governance and management in thailand.
\newblock \textit{Marine Policy}, \textbf{44}, 107--116.

\bibitem[{Brewer et~al.(2013)Brewer, Cinner, Green and Pressey}]{Brewer:2013}
Brewer, T.~D., Cinner, J.~E., Green, A. and Pressey, R.~L. (2013) Effects of
  human population density and proximity to markets on coral reef fishes
  vulnerable to extinction by fishing.
\newblock \textit{Conservation Biology}, \textbf{27}, 443--452.

\bibitem[{Cafri et~al.(2019)Cafri, Wang, Chan and Austin}]{Cafri:2019}
Cafri, G., Wang, W., Chan, P.~H. and Austin, P.~C. (2019) A review and
  empirical comparison of causal inference methods for clustered observational
  data with application to the evaluation of the effectiveness of medical
  devices.
\newblock \textit{Statistical Methods in Medical Research}, \textbf{28},
  3142--3162.

\bibitem[{Campbell et~al.(2020)Campbell, Darling, Pardede, Ahmadia, Mangubhai,
  Amkieltiela, Estradivari and Maire}]{Cambell:2020}
Campbell, S.~J., Darling, E.~S., Pardede, S., Ahmadia, G., Mangubhai, S.,
  Amkieltiela, Estradivari and Maire, E. (2020) Fishing restrictions and
  remoteness deliver conservation outcomes for indonesia's coral reef
  fisheries.
\newblock \textit{Conservation Letters}, \textbf{13}, e12698.

\bibitem[{Chen(2007)}]{Chen:2007}
Chen, X. (2007) Chapter 76 large sample sieve estimation of semi-nonparametric
  models.
\newblock In \textit{Handbook of Econometrics}, vol.~6, 5549--5632. Elsevier.

\bibitem[{Davison and Hinkley(1997)}]{Davison:1997}
Davison, A.~C. and Hinkley, D.~V. (1997) \textit{Bootstrap Methods and their
  Application}.
\newblock Cambridge Series in Statistical and Probabilistic Mathematics.
  Cambridge University Press.

\bibitem[{Edgar et~al.(2014)Edgar, Stuart-Smith, Willis, Baker, Banks, Barrett,
  Becerro, Bernard, Berkhout, Buxton, Campbell, Cooper, M, Edgar, Fosterra,
  Galvan, Irigoyen, Kushner, Moura, Parnell, Shears, Soler, Strain and
  Thomson}]{Edgar:2014}
Edgar, G., Stuart-Smith, R., Willis, TJ~andKininmonth, S., Baker, S., Banks,
  S., Barrett, N., Becerro, M., Bernard, A., Berkhout, J., Buxton, C.,
  Campbell, S., Cooper, A., M, D., Edgar, S., Fosterra, G., Galvan, D.,
  Irigoyen, A., Kushner, D., Moura, R., Parnell, P., Shears, N., Soler, G.,
  Strain, E. and Thomson, R. (2014) Global conservation outcomes depend on
  marine protected areas with five key features.
\newblock \textit{Nature}, \textbf{506}, 216--220.

\bibitem[{Efron(1979)}]{Efron:1979}
Efron, B. (1979) {Bootstrap Methods: Another Look at the Jackknife}.
\newblock \textit{The Annals of Statistics}, \textbf{7}, 1--26.

\bibitem[{Ferraro et~al.(2019)Ferraro, Sanchirico and Smith}]{Ferraro:2019}
Ferraro, P.~J., Sanchirico, J.~N. and Smith, M.~D. (2019) Causal inference in
  coupled human and natural systems.
\newblock \textit{Proceedings of the National Academy of Sciences of the United
  States of America}, \textbf{116}, 5311--5318.

\bibitem[{Gill et~al.(2017)Gill, Mascia, Ahmadia, Glew, Lester, Barnes,
  Craigie, Darling, Free, Geldmann, Holst, Jensen, White, Basurto, Coad, Gates,
  Guannel, Mumby, Thomas, Whitmee, Woodley and Fox}]{Gill:2017}
Gill, D.~A., Mascia, M.~B., Ahmadia, G.~N., Glew, L., Lester, S.~E., Barnes,
  M., Craigie, I., Darling, E.~S., Free, C.~M., Geldmann, J., Holst, S.,
  Jensen, O.~P., White, A.~T., Basurto, X., Coad, L., Gates, R.~D., Guannel,
  G., Mumby, P.~J., Thomas, H., Whitmee, S., Woodley, S. and Fox, H.~E. (2017)
  Capacity shortfalls hinder the performance of marine protected areas
  globally.
\newblock \textit{Nature}, \textbf{543}, 665--669.

\bibitem[{Grorud-Colvert et~al.(2021)Grorud-Colvert, Sullivan-Stack, Roberts,
  Constant, e~Costa, Pike, Kingston, Laffoley, Sala, Claudet, Friedlander,
  Gill, Lester, Day, Gon\c{c}alves, Ahmadia, Rand, Villagomez, Ban, Gurney,
  Spalding, Bennett, Briggs, Morgan, Moffitt, Deguignet, Pikitch, Darling,
  Jessen, Hameed, Carlo, Guidetti, Harris, Torre, Kizilkaya, Agardy, Cury,
  Shah, Sack, Cao, Fernandez and Lubchenco}]{Grorud:2021}
Grorud-Colvert, K., Sullivan-Stack, J., Roberts, C., Constant, V., e~Costa,
  B.~H., Pike, E.~P., Kingston, N., Laffoley, D., Sala, E., Claudet, J.,
  Friedlander, A.~M., Gill, D.~A., Lester, S.~E., Day, J.~C., Gon\c{c}alves,
  E.~J., Ahmadia, G.~N., Rand, M., Villagomez, A., Ban, N.~C., Gurney, G.~G.,
  Spalding, A.~K., Bennett, N.~J., Briggs, J., Morgan, L.~E., Moffitt, R.,
  Deguignet, M., Pikitch, E.~K., Darling, E.~S., Jessen, S., Hameed, S.~O.,
  Carlo, G.~D., Guidetti, P., Harris, J.~M., Torre, J., Kizilkaya, Z., Agardy,
  T., Cury, P., Shah, N.~J., Sack, K., Cao, L., Fernandez, M. and Lubchenco, J.
  (2021) The mpa guide: A framework to achieve global goals for the ocean.
\newblock \textit{Science}, \textbf{373}, eabf0861.

\bibitem[{Holland(1986)}]{Holland:1986}
Holland, P.~W. (1986) Statistics and causal inference.
\newblock \textit{Journal of the American Statistical Association},
  \textbf{81}, 945--960.

\bibitem[{Hong and Raudenbush(2006)}]{Hong:2006}
Hong, G. and Raudenbush, S.~W. (2006) Evaluating kindergarten retention policy:
  A case study of causal inference for multilevel observational data.
\newblock \textit{Journal of the American Statistical Association},
  \textbf{101}, 901--910.

\bibitem[{{Horta e Costa} et~al.(2016){Horta e Costa}, Claudet, Franco, Erzini,
  Caro and Gon\c{c}alves}]{Barbara:2016}
{Horta e Costa}, B., Claudet, J., Franco, G., Erzini, K., Caro, A. and
  Gon\c{c}alves, E.~J. (2016) A regulation-based classification system for
  marine protected areas (mpas).
\newblock \textit{Marine Policy}, \textbf{72}, 192--198.

\bibitem[{Huber et~al.(2016)Huber, Camponovo, Bodory and Lechner}]{Huber2016}
Huber, M., Camponovo, L., Bodory, H. and Lechner, M. (2016) A wild bootstrap
  algorithm for propensity score matching estimators.
\newblock \textit{FSES Working Papers 470}, Faculty of Economics and Social
  Sciences, University of Freiburg/Fribourg Switzerland.

\bibitem[{Kamat(2014)}]{Kamat:2014}
Kamat, V. (2014) {"The Ocean is our Farm": Marine Conservation, Food
  Insecurity, and Social Suffering in Southeastern Tanzania}.
\newblock \textit{Human Organization}, \textbf{73}, 289--298.

\bibitem[{Karr et~al.(2015)Karr, Fujita, Halpern, Kappel, Crowder, Selkoe,
  Alcolado and Rader}]{Karr:2015}
Karr, K.~A., Fujita, R., Halpern, B.~S., Kappel, C.~V., Crowder, L., Selkoe,
  K.~A., Alcolado, P.~M. and Rader, D. (2015) Thresholds in caribbean coral
  reefs: implications for ecosystem-based fishery management.
\newblock \textit{Journal of Applied Ecology}, \textbf{52}, 402--412.

\bibitem[{King and Nielsen(2019)}]{King:2019}
King, G. and Nielsen, R. (2019) Why propensity scores should not be used for
  matching.
\newblock \textit{Political Analysis}, \textbf{27}, 435--454.

\bibitem[{Li et~al.(2013)Li, Zaslavsky and Landrum}]{Li:2013}
Li, F., Zaslavsky, A.~M. and Landrum, M.~B. (2013) Propensity score weighting
  with multilevel data.
\newblock \textit{Statistics in Medicine}, \textbf{101}, 3373--3387.

\bibitem[{Otsu and Rai(2017)}]{Otsu:2017}
Otsu, T. and Rai, Y. (2017) Bootstrap inference of matching estimators for
  average treatment effects.
\newblock \textit{Journal of the American Statistical Association},
  \textbf{112}, 1720--1732.

\bibitem[{Pynegar et~al.(2021)Pynegar, Gibbons, Asquith and
  Jones}]{Pynegar:2021}
Pynegar, E.~L., Gibbons, J.~M., Asquith, N.~M. and Jones, J. P.~G. (2021) What
  role should randomized control trials play in providing the evidence base for
  conservation?
\newblock \textit{Oryx}, \textbf{55}, 235–244.

\bibitem[{Rosenbaum and Rubin(1983)}]{Rosenbaum:1983}
Rosenbaum, P.~R. and Rubin, D.~B. (1983) The central role of the propensity
  score in observational studies for causal effects.
\newblock \textit{Biometrika}, \textbf{70}, 41--55.

\bibitem[{Rubin(1980)}]{Rubin:1980}
Rubin, D.~B. (1980) Randomization analysis of experimental data: The fisher
  randomization test comment.
\newblock \textit{Journal of the American Statistical Association},
  \textbf{75}, 591--593.

\bibitem[{Schoenberg(1946)}]{Schoenberg1946}
Schoenberg, I.~J. (1946) Contributions to the problem of approximation of
  equidistant data by analytic functions: Part a.—on the problem of smoothing
  or graduation. a first class of analytic approximation formulae.
\newblock \textit{Quarterly of Applied Mathematics}, \textbf{4}, 45--99.

\bibitem[{Stuart(2007)}]{Stuart:2007}
Stuart, E.~A. (2007) Estimating causal effects using school-level data sets.
\newblock \textit{Educational Researcher}, \textbf{36}, 187--198.

\bibitem[{Stuart(2010)}]{Stuart:2010}
--- (2010) Matching methods for causal inference: A review and a look forward.
\newblock \textit{Statistical science : a review journal of the Institute of
  Mathematical Statistics}, \textbf{25}, 1--21.

\bibitem[{Tibshirani(1996)}]{Tibshirani:1996}
Tibshirani, R. (1996) Regression shrinkage and selection via the lasso.
\newblock \textit{Journal of the Royal Statistical Society. Series B
  (Methodological)}, \textbf{58}, 267--288.

\bibitem[{Toth et~al.(2014)Toth, van Woesik, Murdoch, Smith, Ogden, Precht and
  Aronson}]{Toth:2014}
Toth, L.~T., van Woesik, R., Murdoch, T. J.~T., Smith, S.~R., Ogden, J.,
  Precht, W.~F. and Aronson, R.~B. (2014) Do no-take reserves benefit
  florida’s corals? 14 years of change and stasis in the florida keys
  national marine sanctuary.
\newblock \textit{Coral Reefs}, \textbf{33}, 565--577.

\bibitem[{UNEP-WCMC et~al.(2021)UNEP-WCMC, IUCN and NGS}]{UNEP:2021}
UNEP-WCMC, IUCN and NGS (2021) Protected planet live report 2021.
\newblock \textit{Tech. rep.}, Cambridge UK; Gland, Switzerland; and
  Washington, D.C., USA.

\bibitem[{VanderWeele(2008)}]{VanderWeele:2008}
VanderWeele, T.~J. (2008) Ignorability and stability assumptions in
  neighborhood effects research.
\newblock \textit{Statistics in Medicine}, \textbf{27}, 1934--1943.

\bibitem[{Yang(2018)}]{Yang:2018}
Yang, S. (2018) Propensity score weighting for causal inference with clustered
  data.
\newblock \textit{Journal of Causal Inference}, \textbf{6}.

\bibitem[{Yang and Lok(2018)}]{Yang:2018b}
Yang, S. and Lok, J.~J. (2018) Sensitivity analysis for unmeasured confounding
  in coarse structural nested mean models.
\newblock \textit{Statistica Sinica}, \textbf{28}, 1703--1723.

\bibitem[{Yang and Zhang(2020)}]{Yang:2020}
Yang, S. and Zhang, Y. (2020) Multiply robust matching estimators of average
  and quantile treatment effects.

\end{thebibliography}


\begin{thebibliography}{4}
\expandafter\ifx\csname natexlab\endcsname\relax\def\natexlab#1{#1}\fi
\expandafter\ifx\csname url\endcsname\relax
  \def\url#1{\texttt{#1}}\fi
\expandafter\ifx\csname urlprefix\endcsname\relax\def\urlprefix{URL}\fi

\bibitem[{Abadie and Imbens(2006)}]{Abadie:2006}
Abadie, A. and Imbens, G.~W. (2006) Large sample properties of matching
  estimators for average treatment effects.
\newblock \textit{Econometrica}, \textbf{74}, 235--267.

\bibitem[{Abadie and Imbens(2011)}]{Abadie:2011}
--- (2011) Bias-corrected matching estimators for average treatment effects.
\newblock \textit{Journal of Business \& Economic Statistics}, \textbf{29},
  1--11.

\bibitem[{Otsu and Rai(2017)}]{Otsu:2017}
Otsu, T. and Rai, Y. (2017) Bootstrap inference of matching estimators for
  average treatment effects.
\newblock \textit{Journal of the American Statistical Association},
  \textbf{112}, 1720--1732.

\bibitem[{Serfling(1968)}]{Serfling:1968}
Serfling, R.~J. (1968) Contributions to central limit theory for dependent
  variables.
\newblock \textit{The Annals of Mathematical Statistics}, \textbf{39},
  1158--1175.

\end{thebibliography}

\makeatletter\@input{sup_main.tex}\makeatother
\end{document}


\def\spacingset#1{\renewcommand{\baselinestretch}%
{#1}\small\normalsize} \spacingset{1}

\maketitle

\singlespacing

Section \ref{s:app} and \ref{s:app_asymp} provide the proofs of Theorem \ref{thm:ate}, \ref{thm:att} and \ref{thm:asymp_valid}. Section \ref{s:cluster_boot_att} describes the proposed cluster-weighted bootstrap for the average treatment effect on the treated (ATT). Section \ref{s:sup_figure} presents additional figures from the analysis of conservation policy effects on marine biodiversity.

\section{Proofs of Theorem \ref{thm:ate} and \ref{thm:att}} \label{s:app}
Let $S$ be a generic variable for matching which could be at cluster-level or unit-level. The original estimator of average treatment effect (ATE) is
$$
\hat{\tau}_{\text{mat}} = N^{-1}\sum^N_{i=1}\sum^R_{r=1}(2A_{ir} - 1)\left\{1+\frac{1}{M}K_{M}(i,r)\right\}Y_{ir}.
$$
According to \cite{Abadie:2006}, we write $\sqrt{N}(\hat{\tau}_{\text{mat}} - B_M - \tau) = \sqrt{N}(\overline{\tau(S)} - \tau) + \sqrt{N}E_M$, where 
\begin{align*}
    \overline{\tau(S)} - \tau &= \frac{1}{N}\sum^N_{i=1}\sum^R_{r=1}\mu_1(S_{ir}) - \mu_0(S_{ir}) - \tau,\\
    E_M &= \frac{1}{N}\sum^N_{i=1}E_{M,i}=\frac{1}{N}\sum^N_{i=1}\sum^R_{r=1}(2A_{r} - 1)\left\{1+\frac{K_M(i,r)}{M}\right\}\left\{Y_{ir} - \mu_{A_{r}}(S_{ir})\right\}.
\end{align*}
Then for the first part $\sqrt{n}(\overline{\tau(S)} - \tau)$, by a standard central limit theorem, 
$$
\sqrt{N}(\overline{\tau(S)} - \tau) \stackrel{d}{\rightarrow} N(0, V^{\tau(S)}).
$$

Consider the distribution of $\sqrt{N}E_M/\sqrt{V^E}$, we adopt the results shown in \cite{Abadie:2006} that the moments of $K_M(i,r)$ are bounded uniformly in $N$. For a given $S, A$, the Lindeberg-Feller condition requires that 
$$
\frac{1}{NV^E}\sum^N_{i=1}\mathbb{E}\left[(E_{M,i}^2) \mathbb{I}\left\{|E_{M,i}| \geq \eta\sqrt{NV^E}\right\}|S,A\right] \rightarrow 0
$$
for all $\eta > 0$.
By the same routine, this condition holds by using H\"older's and Markov's inequalities,
\begin{align*}
    \mathbb{E}\left[(E_{M,i}^2) \mathbb{I}\left\{|E_{M,i}| \geq \eta\sqrt{NV^E}\right\}|X,A\right] \\
    \leq \mathbb{E}\left[(E_{M,i}^4|S, A)\right)^{1/2}\left(\mathbb{E}\left[\mathbb{I}\left\{|E_{M,i}| \geq \eta\sqrt{NV^E}\right\}|S,A\right]\right)^{1/2}\\
    \leq \mathbb{E}\left[(E_{M,i}^4|S, A)\right)^{1/2} \left(\mbox{Pr}(|E_{M,i}| \geq \eta\sqrt{NV^E}|S, A)\right)\\
    \leq \mathbb{E}\left[(E_{M,i}^4|S, A)\right)^{1/2} \frac{\mathbb{E}[(E_{M,i})^2|S,A]}{\eta^2NV^E}.
\end{align*}
Let $\overline{\sigma}^2 = \mbox{sup}_{a,s}\sigma^2(s,a) < \infty, \underline{\sigma}^2 = \mbox{inf}_{a,s}\sigma^2(s,a) > 0$ and $\overline{C}=\mbox{sup}_{a,s}\mathbb{E}[\left\{Y_{ir} - \mu_{A_{r}}(S_{ir})\right\}^4|S_{ir}=s, A_{r}=a] < \infty$, then the condition is bounded as
$$
\frac{1}{NV^E}\sum^N_{i=1}\mathbb{E}\left[(E_{M,i}^2) \mathbb{I}\left\{|E_{M,i}| \geq \eta\sqrt{NV^E}\right\}|S,A\right]  \leq \frac{\overline{\sigma}^2\overline{C}^{1/2}}{\eta^2\underline{\sigma}^4N}(\frac{1}{N}\sum^N_{i=1}(\sum^R_{r=1}(1+M^{-1}K_M(i,r)))^4).
$$
By central limit theorem for dependent variables \citep{Serfling:1968}, with finite $V^E$, we have
$$
\sqrt{N}E_M \stackrel{d}{\rightarrow} N(0, V^{E}).
$$
Therefore, the matching estimator $\hat{\tau}_{\text{mat}}$ follows asymptotic normal distribution with finite variance $(V^E + V^{\tau(S)})/N$, after ignoring the conditional bias term. 

For the estimator of average treatment effect on the treated (ATT), 
$$
\hat{\tau}_{\text{mat}}^t = \frac{1}{N_1}\sum^N_{i=1}\sum^R_{r=1}\left\{A_{r} - (1-A_{r})\frac{K_M(i,r)}{M}\right\}Y_{ir},
$$
where $N_1$ is the number of units in treatment group. After ignoring the conditional bias term $B_M^t$, 
\begin{align*}
    \tilde{\tau}^t &= \overline{\tau(X)}^t - B_M^t\\
    &= \frac{1}{N_1}\sum^N_{i=1}\sum^R_{r=1}A_{ir}\left\{\mu(S_{ir},1) - \mu_0(S_{ir})\right\} \\
    &+ \frac{1}{N_1}\sum^N_{i=1}\sum^R_{r=1}\left\{A_{r} - (1-A_{r})\frac{K_M(i,r)}{M}\right\}\left\{Y_{ir} - \mu_{A_{r}}(S_{ir})\right\}\\
    &= \frac{1}{N_1}\left[\sum^R_{r=1}A_{r}\left\{\mu(S_{ir},1) - \mu_0(S_{ir})\right\} + \left\{A_{r} - (1-A_{r})\frac{K_M(i,r)}{M}\right\}\left\{Y_{ir} - \mu_{A_{r}}(S_{ir})\right\} \right]\\
    &= \overline{\tau(S)}^t + E_M^t,
\end{align*}
where 
\begin{align*}
    \overline{\tau(S)}^t &= \frac{1}{N_1}\sum^N_{i=1}\sum^R_{r=1}A_{r}\left\{\mu(S_{ir},1) - \mu_0(S_{ir})\right\},\\
    E_M^t &= \frac{1}{N_1}\sum^N_{i=1}\sum^R_{r=1}\left\{A_{r} - (1-A_{r})\frac{K_M(i,r)}{M}\right\}\left\{Y_{ir} - \mu_{A_{r}}(S_{ir})\right\}.
\end{align*}

Consider $\sqrt{N_1}(\hat{\tau}^t_{\text{mat}} - B_M^t - \tau^t) = \sqrt{N_1}(\overline{\tau(S)}^t - \tau^t) + \sqrt{N_1}E_M^t$, according to the similar procedure described above, the first part 
$\sqrt{N_1}(\overline{\tau(S)}^t - \tau^t) \stackrel{d}{\rightarrow} N(0,V^{\tau(S),t})$
by the standard central limit theorem. The second part includes $$E_{M}^t = \sum^R_{r=1}\left\{A_{r} - (1-A_{r})\frac{K_M(i,r)}{M}\right\}\left\{Y_{ir} - \mu_{A_{r}}(S_{ir})\right\}.$$ By the central limit theorem for dependent variables, we have
$$
\sqrt{N}E_M^t \stackrel{d}{\rightarrow} N(0, V^{E,t}).
$$
The matching estimator $\hat{\tau}_{\text{mat}}^t$ thus follows an asymptotic normal distribution with finite variance $(V^{E,t} + V^{\tau(S),t})/N_1$, after ignoring the conditional bias term. 

Under certain regularity conditions in \citep{Abadie:2011} and the condition for outcome mean functions \citep{Otsu:2017}, we have
$\sqrt{N}(\hat{B}_M - B_M) \stackrel{p}{\rightarrow} 0$ and $\sqrt{N}(\hat{B}_M^t - B_M^t) \stackrel{p}{\rightarrow} 0$. Therefore, the above asymptotic normality and the fast convergence of estimated condition bias terms help complete the proofs for Theorem 1 and Theorem 2.

\section{Proof of Theorem \ref{thm:asymp_valid}} \label{s:app_asymp}
For simplicity of notation, we focus on the case with balanced cluster size and reconstruct the matching estimator for convenience. Denote $N=Rn$ as the number of observations, $R$ as the number of balanced clusters, $n$ as the cluster size for the each cluster, and $S$ as a unified variable that represents either the cluster-level covariates or stacked covariates, the matching estimator is
$$
\hat{\tau}_{\text{mat}} = \frac{1}{Rn}\sum^R_{r=1}\sum^{n}_{j=1}(2A_r - 1)\{1 + M^{-1}K_M(r,j)\}Y_{rj}.
$$
The matching estimator can be decomposed into three parts, $\hat{\tau}_{\text{mat}} = \overline{\tau(S)} + E_M + B_M$,
where 
$$\overline{\tau(X,Z)} = \frac{1}{Rn}\sum^R_{r=1}\sum^n_{j=1}\left\{\mu_1(S_{rj}) - \mu_0(S_{rj})\right\},
$$
$$
E_M = \frac{1}{Rn}\sum^R_{r=1}\sum^n_{j=1}(2A_{r} - 1)\left\{1+\frac{K_M(r,j)}{M}\right\}\left\{Y_{rj}-\mu_{A_{r}}(S_{rj})\right\},
$$
$$B_M = \frac{1}{Rn}\sum^R_{r=1}\sum^n_{j=1}(2A_{r} - 1) \left[\frac{1}{M}\sum_{(k,l)\in \mathcal{J}_M(r,j)}\left\{\mu_{1 - A_{r}}(S_{rj}) - \mu_{1-A_{r}}(S_{kl})\right\}\right].
$$

The debiased estimator for cluster weighted bootstrap is expressed as
\begin{align*}
    \tilde{\tau} &= \hat{\tau}_{\text{mat}} - \hat{B}_M\\
    &= \frac{1}{Rn}\sum^R_{r=1}\sum^{n}_{j=1}\{\hat{\mu}_1(S_{rj}) - \hat{\mu}_0(S_{rj})\} \\
    &+ \frac{1}{Rn}\sum^R_{r=1}\sum^{n}_{j=1}(2A_r - 1)\{1 + M^{-1}K_M(r,j)\}\{Y_{rj} - \hat{\mu}_{A_r}(S_{rj})\}\\
    &= \frac{1}{Rn}\sum^R_{r=1}\sum^{n}_{j=1}\left[\{\hat{\mu}_1(S_{rj}) - \hat{\mu}_0(S_{rj})\} + (2A_r - 1)\{1 + M^{-1}K_M(r,j)\}\{Y_{rj} - \hat{\mu}_{A_r}(S_{rj})\}\right]\\
    &= \frac{1}{R}\sum^R_{r=1}\tilde{\tau}_r,
\end{align*}
where $\tilde{\tau}_r = \frac{1}{n}\sum^{n}_{j=1}\left[\{\hat{\mu}_1(S_{rj}) - \hat{\mu}_0(S_{rj})\} + (2A_r - 1)\{1 + M^{-1}K_M(r,j)\}\{Y_{rj} - \hat{\mu}_{A_r}(S_{rj})\}\right]$.

Let $\hat{e}_{rj} = Y_{rj} - \hat{\mu}_{A_r}(S_{rj}), \hat{\xi}_{rj} = (2A_r - 1)\{\hat{\mu}(A_r, S_{rj}) - \hat{\mu}(1-A_r, S_{rj})\} - \tilde{\tau}$, then
$$
\tilde{\tau}_r - \tilde{\tau} = \frac{1}{n}\sum^{n}_{j=1}\left[(2A_r - 1)\{1 + M^{-1}K_M(r,j)\}\hat{e}_{rj} + \hat{\xi}_{rj}\right].
$$
To show that the cluster weighted bootstrap is a valid approach for statistical inference, we need to show the consistency of cluster weighted bootstrap variance, i.e., 
$$E\left[\left\{\sum^R_{r=1}W_r^*(\tilde{\tau}_r - \tilde{\tau})|(\bm{Y},\bm{A},\bm{S})\right\}^2\right] \stackrel{p}{\rightarrow} \sigma^2,$$
where $W_r^*=M_r^*/\sqrt{R}$, and $(M_1^*, \ldots, M_R^*)$ is required to satisfy Assumption \ref{a:weight}, e.g., it could be a vector from a multinomial distribution with equal probability. The proof follows the same strategy as \citet{Otsu:2017}, adjusting for the clustered case. 

\begin{assumption}\label{a:weight}
(Conditions for $W_r^*$) 
\begin{enumerate}[label=(\roman*)]
    \item $(W_1^*, \ldots, W_R^*)$ is exchangeable and independent of $\bm{Z}=(\bm{Y},\bm{A},\bm{S})$.
    \item $\sum^R_{r=1}(W_r^* - \bar{W}^*)^2 \stackrel{p}{\rightarrow} 1$ where $\bar{W}^* = R^{-1}\sum^R_{r=1}W_r^*$.
    \item $\mbox{max}_{1,\ldots,R}\lvert W_r^* - \bar{W}^* \rvert \stackrel{p}{\rightarrow} 0$.
    \item $E[W_r^{*2}] = O(R^{-1})$ for all $r=1,\ldots, R$.
\end{enumerate}
\end{assumption}

Consider $ \sqrt{R}T^* = \sum^R_{r=1}(W_r^* - \bar{W}^*)(\tilde{\tau}_r - \tilde{\tau})$, it can be decomposed into three parts, 
\begin{align*}
    \sqrt{R}T^* &= \sum^R_{r=1}(W_r^* - \bar{W}^*)(\tilde{\tau}_r - \tilde{\tau})\\
    &= \sum^R_{r=1}(W_r^* - \bar{W}^*)\left[\frac{1}{n}\sum^{n}_{j=1}\left\{(2A_r - 1)\{1 + M^{-1}K_M(r,j)\}\hat{e}_{rj} + \hat{\xi}_{rj}\right\}\right]\\
    &= \sqrt{R}(T_N^* + R_{1N}^* + R_{2N}^*),
\end{align*}
where
\begin{align*}
    \sqrt{R}T_N^* &= \sum^R_{r=1}(W_r^* - \bar{W}^*)\left[\frac{1}{n}\sum^{n}_{j=1}\left\{(2A_r - 1)\{1 + M^{-1}K_M(r,j)\}e_{rj} + \xi_{rj}\right\}\right],\\
    \sqrt{R}R_{1N}^* &= \sum^R_{r=1}(W_r^* - \bar{W}^*)\left[\frac{1}{n}\sum^{n}_{j=1}(2A_r - 1)\{1 + M^{-1}K_M(r,j)\}\{\mu(A_r, S_{rj}) - \hat{\mu}(A_r,S_{rj})\}\right],\\
    \sqrt{R}R_{2N}^* &= \sum^R_{r=1}(W_r^* - \bar{W}^*)\left[\frac{1}{n}\sum^{n}_{j=1}(\hat{\xi}_{rj} - \xi_{rj})\right].
\end{align*}
We need to show that $\mbox{Pr}\{\sqrt{R}\lvert R_{1N}^* \rvert > \epsilon\} \stackrel{p}{\rightarrow} 0$, $\mbox{Pr}\{\sqrt{R}\lvert R_{2N}^* \rvert > \epsilon\} \stackrel{p}{\rightarrow} 0$, and $E\left[(\sqrt{R}T_N^*)^2\right]  \stackrel{p}{\rightarrow} \sigma^2$. 

For $\sqrt{R}R_{1N}^*$, the Markov's inequality is leveraged to show its convergence toward 0 as $R \rightarrow \infty$. 
\begin{align*}
    E[(\sqrt{R}R_{1N}^*)^2] &= 
    E\left[\left[\sum^R_{r=1}(W_r^* - \bar{W}^*)\gamma_r\right]^2\right]\\
    &= RE[(W_1^* - \bar{W}^*)^2]\frac{1}{R}\sum^R_{r=1}\gamma_r^2 \\
    &+ R(R-1)E\left[(W_1^* - \bar{W}^*)(W_2^* - \bar{W}^*)\right] \times \frac{1}{R(R-1)}\sum_{i\neq k}\gamma_r\gamma_k
\end{align*}
where $\gamma_r = \frac{1}{n}\sum^{n}_{j=1}(2A_r - 1)\{1 + M^{-1}K_M(r,j)\}\{\mu(A_r, S_{rj}) - \hat{\mu}(A_r,S_{rj})\}$. Since $RE[(W_1^* - \bar{W}^*)^2] = O(1), R(R-1)E\left[(W_1^* - \bar{W}^*)(W_2^* - \bar{W}^*)\right] = O(1),$ $\lvert\mu(a,\cdot) - \hat{\mu}(a,\cdot)\rvert_{k-1}=o_p(N^{-1/2 + 1/k})$, and $E(K_M(i,j)^q)$ is uniformly bounded over $N$, we obtain $E[(\sqrt{R}R_{1N}^*)^2] \stackrel{p}{\rightarrow} 0$. Then the Markov inequality shows that $\mbox{Pr}[\sqrt{R}\lvert R_{1N}^* \rvert > \epsilon]\stackrel{p}{\rightarrow} 0$.

Similarly, $\mbox{Pr}[\sqrt{R}\lvert R_{2N}^* \rvert > \epsilon]\stackrel{p}{\rightarrow} 0$ by Markov inequality, since
\begin{align*}
    E\left[(\sqrt{R}R^*_{2N})^2\right] &= E\left[\left[\sum^R_{r=1}(W_r^* - \bar{W}^*)\{\frac{1}{n}\sum^{n}_{j=1}(\hat{\xi}_{rj} - \xi_{rj})\}\right]^2\right]\\
    &= RE[(W_1^* - \bar{W}^*)^2]\frac{1}{R}\sum^R_{r=1}\{\frac{1}{n}\sum^{n}_{j=1}(\hat{\xi}_{rj} - \xi_{rj})\}^2 \\
    &+ R(R-1)E\left[(W_1^* - \bar{W}^*)(W_2^* - \bar{W}^*)\right] \\
    &\times \frac{1}{R(R-1)}\sum_{i\neq k}\{\frac{1}{n}\sum^{n}_{j=1}(\hat{\xi}_{rj} - \xi_{rj})\}\{\frac{1}{n}\sum^{n}_{l=1}(\hat{\xi}_{kl} - \xi_{kl})\}\\
    &\stackrel{p}{\rightarrow} 0.
\end{align*}

For $\sqrt{R}T_N^*$, we shall show $E[(\sqrt{R}T_N^*)^2 \mid \bm{Z}] = \sigma^2$.
\begin{align*}
    E[(\sqrt{R}T_N^*)^2\mid \bm{Z}] &= E\left[\left[\sum^R_{r=1}(W_r^* - \bar{W}^*)\frac{1}{n}\sum^{n}_{j=1}\left\{(2A_r - 1)\{1 + M^{-1}K_M(r,j)\}e_{rj} + \xi_{rj}\right\}\right]^2\right]\\
    &= RE[(W_1^*-\bar{W}^*)^2]\\
    &\times\frac{1}{R}\sum^R_{r=1}E\left[[\frac{1}{n}\sum^{n}_{j=1}\left\{(2A_r - 1)\{1 + M^{-1}K_M(r,j)\}e_{rj} + \xi_{rj}\right\}]^2\right] \\
    & + R(R-1)\sum_{r,j \neq k,l}E[(W_1^*-\bar{W}^*)(W_2^*-\bar{W}^*)]\\
    &\times \frac{1}{R(R-1)}E[\frac{1}{n}\sum^{n}_{j=1}\left\{(2A_r - 1)\{1 + M^{-1}K_M(r,j)\}e_{rj} + \xi_{rj}\right\}]\\
    &\times E[\frac{1}{n}\sum^{n}_{k=1}\left\{(2A_k - 1)\{1 + M^{-1}K_M(k,l)\}e_{kl} + \xi_{kl}\right\}]\\
    & \stackrel{p}{\rightarrow} \sigma^2
\end{align*}
since $RE[(W_1^* - \bar{W}^*)^2] = O(1), R(R-1)E\left[(W_1^* - \bar{W}^*)(W_2^* - \bar{W}^*)\right] = O(1),$ and $\sigma^2$ is the variance of the debiased matching estimator $\tilde{\tau}$. Therefore, it is enough to show $  E\left[\left\{\sum^R_{r=1}W_r^*(\tilde{\tau}_r - \tilde{\tau})\mid \bm{Z}\right\}^2\right] \stackrel{p}{\rightarrow} \sigma^2$.


\section{Cluster Weighted Bootstrap for the Average Treatment Effect on the Treated}\label{s:cluster_boot_att}
According to Theorem \ref{thm:att}, under certain conditions, the bias-corrected estimator for the average treatment effect on the treated is asymptotically normal, i.e.,
\[
\frac{\sqrt{N_1}(\hat{\tau}^t  - \tau^t)}{\sigma^t} \stackrel{d}{\rightarrow} N(0,1),
\]
where
\begin{align*}
    (\sigma^t)^2 &= (\sigma_{1}^t)^2 + (\sigma_2^t)^2,\\
    (\sigma_{1}^t)^2 &= \frac{1}{N_1}\sum^N_{i=1}\sum^R_{r=1}\left\{A_{r} - (1-A_{r})M^{-1}K_M(i,r)\right\}^2\sigma^2(X_{ir}, Z_r, A_{r}),\\
    (\sigma_2^t)^2 &= E\left[\left\{(\mu(X_{ir},Z_r, 1) - \mu(X_{ir},Z_r, 0))-\tau^t\right\}^2|A_{r}=1\right].
\end{align*}

With bias correction, we rewrite our bias-corrected estimator $\hat{\tau}^t = \hat{\tau}_{\text{mat}}^t - \hat{B}_M^t$,
\begin{align*}
    \hat{\tau}^t &= \frac{1}{N_1}\sum^N_{i=1}\sum^R_{r=1}\left[A_{r}\left\{Y_{ir} - \hat{\mu}(X_{ir},Z_r, 1-A_{r}\right\}\right] + (1-A_{r})\frac{K_M(i,r)}{M}\left\{Y_{ir} - \hat{\mu}( X_{ir}, Z_r, A_{r})\right\}\\
    &= \frac{1}{N_1}\sum^N_{i=1}\hat{\tau}_i^t
\end{align*}
Then with suitable generated weights $\{W_i^*\}_{i=1}^N$, 
\[
\sqrt{N_1}(\hat{\tau}^t - \tau^t) = \sum^N_{i=1}W_i^*(\hat{\tau}_i^t - \hat{\tau}^t)  \stackrel{d}{\rightarrow} N(0,(\sigma^t)^2).
\]
Applying to our clustered data, we proposed the following cluster weighted bootstrap method on variance estimation for the average treatment effect on the treated.
\begin{itemize}
    \item \textit{\textbf{Step 1'}}: Obtain the weighted bootstrap samples $\{\hat{\tau}_i\}^N_{i=1}$ based on the matching estimator framework.
    \item \textit{\textbf{Step 2'}}: For clustered data with $N$ observations and $R$ non-overlapped clusters, sample $R$ clusters with replacement.
    \item \textit{\textbf{Step 3'}}: Include all $\{\hat{\tau}_i\}^N_{i=1}$ within selected clusters and calculate their corresponding weights $\{W_i^*\}^N_{i=1}$. One option of generating weights is to set $W_i^*=M_i^*/\sqrt{N}$, where $(M_1^*, \ldots, M_N^*)$ is a vector from a multinomial distribution with equal probability. 
    \item \textit{\textbf{Step 4'}}: Obtain a bootstrap replicate as $\hat{\tau}^*_b = \sum^N_{i=1}W_i^*(\hat{\tau}_i - \hat{\tau})$.
    \item \textit{\textbf{Step 5'}}: Repeat the Step 1'-4' for $B$ times. Compute the bootstrap variance estimator for the bias-corrected matching estimator $\hat{\tau}$ as the empirical variance of $\{\hat{\tau}^*_b\}^B_{b=1}$.
\end{itemize}

\section{Supplementary Figures} \label{s:sup_figure}
\begin{figure}[H]
\centering
\begin{subfigure}{0.75\textwidth}
\centering
\includegraphics[width=\textwidth]{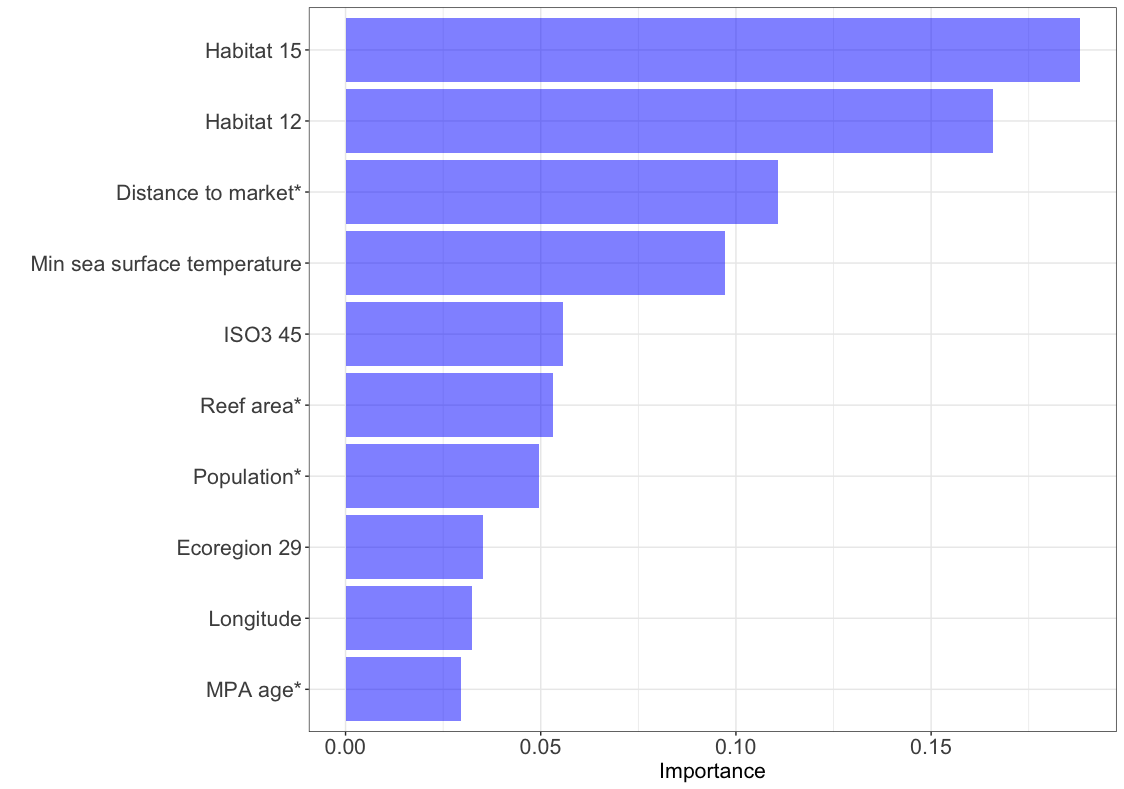}
\caption{Multi-use}
\end{subfigure}
\medskip
\begin{subfigure}{0.75\textwidth}
\centering
\includegraphics[width=\textwidth]{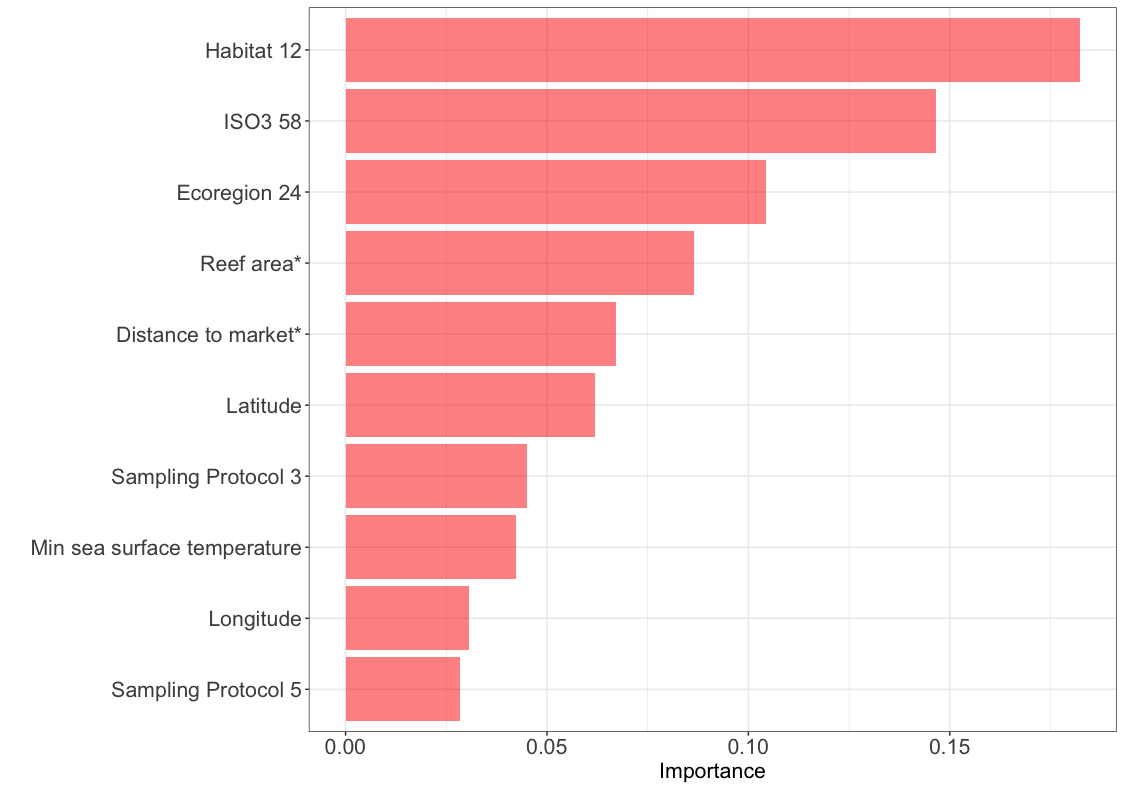}
\caption{No-take}
\end{subfigure}
\caption{Top 10 important covariates under different marine protected area policies by regression forest (*indicates Box-Cox transformation for the covariate).}
\label{f:imp_plot}
\end{figure}

\bibliographystyle{rss.bst}
\bibliography{supplementary.bib}
\makeatletter\@input{match_sup.tex}\makeatother